\documentclass[conference]{IEEEtran}
\IEEEoverridecommandlockouts
\usepackage{cite}
\usepackage{amsmath,amssymb,amsfonts}
\usepackage{algorithmic}
\usepackage{graphicx}
\graphicspath{{figures/}}
\usepackage{textcomp}
\usepackage{xcolor}

\usepackage{flushend}

\usepackage{colortbl}

\usepackage{subfigure}

\usepackage{float}
\usepackage{multirow,bigdelim}
\usepackage{xspace}
\usepackage{booktabs}

\usepackage{verbatim}

\usepackage{url}

\usepackage{color}

\usepackage{fancyhdr}

\def\BibTeX{{\rm B\kern-.05em{\sc i\kern-.025em b}\kern-.08em
    T\kern-.1667em\lower.7ex\hbox{E}\kern-.125emX}}
\begin{document}

\title{\huge 
Predicting Crash Fault Residence via Simplified Deep Forest Based on A Reduced Feature Set \\
}

\author{\IEEEauthorblockN{Kunsong Zhao\IEEEauthorrefmark{2},
Jin Liu\thanks{* Corresponding authors: jinliu@whu.edu.cn, zhouxullx@cqu.edu.cn}\IEEEauthorrefmark{2}\IEEEauthorrefmark{1},
Zhou Xu\IEEEauthorrefmark{3}\IEEEauthorrefmark{1}, Li Li\IEEEauthorrefmark{4}, Meng Yan\IEEEauthorrefmark{3}, Jiaojiao Yu\IEEEauthorrefmark{2}, Yuxuan Zhou\IEEEauthorrefmark{5}
}
\IEEEauthorblockA{\IEEEauthorrefmark{2}School of Computer Science, Wuhan University, Wuhan, China}
\IEEEauthorblockA{\IEEEauthorrefmark{3}School of Big Data and Software Engineering, Chongqing University, Chongqing, China}
\IEEEauthorblockA{\IEEEauthorrefmark{4}Faculty of Information Technology, Monash University, Melbourne, Australia}
\IEEEauthorblockA{\IEEEauthorrefmark{5}College of Engineering and Computer Science, Syracuse University, Syracuse, USA}
}

\maketitle

\begin{abstract}
	The software inevitably encounters the crash, which will take developers a large amount of effort to find the fault causing the crash (short for crashing fault). Developing automatic methods to identify the residence of the crashing fault is a crucial activity for software quality assurance. Researchers have proposed methods to predict whether the crashing fault resides in the stack trace based on the features collected from the stack trace and faulty code, aiming at saving the debugging effort for developers. However, previous work usually neglected the feature preprocessing operation towards the crash data and only used traditional classification models. In this paper, we propose a novel crashing fault residence prediction framework, called ConDF, which consists of a consistency based feature subset selection method and a state-of-the-art deep forest model. More specifically, first, the feature selection method is used to obtain an optimal feature subset and reduce the feature dimension by reserving the representative features. 
	Then, a simplified deep forest model is employed to build the classification model on the reduced feature set. The experiments on seven open source software projects show that our ConDF method performs significantly better than 17 baseline methods on three performance indicators.
\end{abstract}


\section{Introduction}
Software has become a popular service, which plays an indispensable part in our daily life. However, as the increase of the impacts of software complexity and many uncertainties, it is inevitable to occur faults during the process of software development. Since the faults in the software can trigger software crashes, it is vital to predict the residence of faults causing the crash (short for crashing faults) and then fix them, which is a hot issue for software quality assurance in recent years \cite{wong2016survey}.

When a program unit crashes, the software automatically generates a crash report which consists of the stack trace to record the exception information of this unit at this time, such as the function invocation information. This kind of information is useful to identify the root cause of the crashing fault. Predicting whether the crashing faults locate in the stack trace or not can help the developers save the debugging effort. For example, if a crashing fault locates inside the stack trace, developers just need to focus on the corresponding code recorded in the stack trace. Otherwise, if a crashing fault locates outside the stack trace, developers have to spend a great amount of effort checking the function invocation graph, which involves in inspecting many lines of code. The goal of the crashing fault residence prediction task is to identify whether the crashing fault locates inside the stack trace or not, expecting to promote the debugging process. 

Researchers have recently begun to focus on the topic of crashing fault residence prediction. Gu et al. \cite{gu2019does} proposed to extract the features from the stack trace and faulty code for this task. They proposed an automatic method called CraTer for this purpose. More specifically, they extracted 89 features from the stack trace and source code to characterize the crash instances. For each crash instance, if the code information of the crashing fault exactly matches the information recorded in the stack trace, this crash instance is deemed as inside the stack trace, otherwise, outside the stack trace. Finally, they applied some traditional machine learning classifiers to conduct the experiments. Following Gu et al.'s work, Xu et al. \cite{xu2020imbalanced} proposed an imbalance learning method with kNN classification, and Xu et al. \cite{xu2019identify} developed a cross project model with logistic regression classifier to predict the residence of crashing faults.

However, the aforementioned work neglected the selection process for representative features and only used traditional classifiers for their purpose.
In general, the prediction performance of a classification task is highly correlated with the feature quality of the crash data and the employed classification methods. Thus, it is crucial to apply feature engineering techniques to preprocess the crash data to obtain the optimal feature subset (or feature representation) and employ effective classification models to achieve satisfactory performance. 
In this work, we propose a novel method that consists of a feature subset selection method and a state-of-the-art classification model for the crashing fault residence prediction task. More specifically, we apply the \textbf{Con}sistency-based feature subset selection method (\textbf{Con}) to reserve a smallest feature subset that has the same discriminability as the original feature set. 
After the feature selection process, a simplified \textbf{D}eep \textbf{F}orest (\textbf{DF}) method is applied to the reduced feature set to build the classification model for the crashing fault residence prediction. DF generates a decision tree ensemble approach with a cascade structure, which enhances the ability of representation learning for high-dimensional features via a multi-grained scanning method \cite{zhou2017deep}. As the crash data after feature reduction only contains low-dimensional features, we apply the simplified version of DF without the multi-grained scanning to build our prediction model to identify the residence of the crashing faults. In this work, we call our crashing fault residence prediction method ConDF as it consists of two main techniques, i.e., \textbf{Con} method for feature subset selection (i.e., feature reduction) and the simplified \textbf{DF} method for classification  model construction.

To evaluate the effectiveness of our proposed ConDF method for the crashing fault residence prediction task, we conduct experiments on seven open source software projects with three indicators. The experimental results show that ConDF achieves average F-measure for crash instances inside the stack trace of 0.722, average F-measure for crash instances outside the stack trace of 0.942, and average MCC of 0.681. In addition, ConDF obtains significantly better performance than 17 baseline methods on all three indicators.

We highlight the main contributions of this paper as follows:
\begin{itemize}
	\item We propose a novel compositional method, called ConDF, that combines a feature subset selection method and an advanced deep forest model for the crashing fault residence prediction task. 
	\item 
	We apply the consistency-based feature subset selection method to pick up the most representative features of the crash data. In addition, we are the first to introduce the deep forest model to predict the residence of the crashing faults. 
	\item We employ crash data from seven open source projects as benchmark dataset and evaluate our ConDF method using three indicators. The results of sufficient empirical evaluations show that the performance of our proposed ConDF method presents significant superiority compared with 17 baseline methods.
\end{itemize}

The remainder of this paper is organized as follows.  Section \ref{sec2} introduces the related work. Section \ref{sec3} describes the detail of our proposed ConDF method. Section \ref{sec4} describes our experimental setup. Section \ref{sec5} reports the experimental results. Section \ref{sec6} describes the potential threats to validity of our work. Finally, Section \ref{sec7} concludes our work and presents future work.

\section{Background and Related Work}\label{sec2}
As the aim of our work is to predict whether the crashing faults locate in the frame of the stack trace or not based on the part of features extracted from stack trace information, we first present the background of stack trace and analyze some previous related studies about crash reproduction and crash localization based on stack traces. As we employ a feature selection technique for data preprocessing and a deep forest for classifier construction in the first and second stage of our method respectively, we present some related work involving in feature selection and deep forest for software engineering tasks individually.

\subsection{Stack Trace Analysis}

When the software crashes, the system automatically throws the exceptions. The stack trace records the information of these exceptions, such as the function invocation sequence related to the exception and the corresponding type, which helps developers analyze where the program goes wrong and reduces the cost of the efforts of developers. The stack trace consists of multiple frame objects, in which the first frame records the type of exception and other frames record the information of function invocation.

Chen et al. \cite{chen2014star} proposed a STAR framework, which extracted the crash information such as the exception types, names, and line numbers from the stack traces and combined a backward symbolic execution with a sequence composition technique to reproduce crashes. Their experiments on three projects showed that STAR successfully exploited 59.6\% of crashes. 
Nayrolles et al. \cite{nayrolles2015jcharming}\cite{nayrolles2017bug} proposed a novel crash reproduction approach, called JCHARMING, which extracted exception information from the stack traces and detected the buggy crashes using model checking. Their experiments on 10 open source software systems showed that JCHARMING was impactful in reproducing bugs from different systems.
Xuan et al. \cite{xuan2015crash} proposed an approach, called MUCRASH, which extracted the classes from the stack traces and reproduced the crashes via test case mutation. Their experiments on 12 crashes showed that MUCRASH reproduced the same stack trace on 7 out of 12 crashes. 
Soltani et al. \cite{soltani2017guided} proposed a post-failure method, called EvoCrash, which used the stack traces to guide the search process of the genetic algorithm to reduce the search space and eliminate limitations of replicating crashes in the real world. Their empirical study on three projects showed that EvoCrash could replicate 82\% of real-world crashes. 
Sabor et al. \cite{sabor2019automatic} proposed a new approach that generated feature vectors from the collected stack traces, and then combined categorical features to predict the bug severity. They also used a cost-sensitive KNN method to release the issue of imbalanced data. Their experiments on Eclipse project showed that this approach could improve the prediction accuracy.
Soltani et al. \cite{soltani2020benchmark} proposed a benchmark of real-world crashes extracted from the stack traces, called JCrashPack, which contains 200 crashes derived from seven Java projects. Their empirical study showed the effectiveness of search-based crash reproduction method on real-world crashes. 

Recently, there are many studies using stack traces for crash localization, which is related to our work. 
Wu et al. \cite{wu2014crashlocator} proposed a method, called CrashLocator, using the information of the static call graph from the stack trace to locate crashes. Their results on real-world Mozilla crash data showed the effectiveness of this method.
Wong et al. \cite{wong2014boosting} developed a new tool, called BRTracer, employing the segmentation of source code files and the stack trace analysis of bug reports to identify buggy files. Their results on Eclipse, AspectJ, and SWT showed that BRTracer could achieve performance improvement of bug localization.
Moreno et al. \cite{moreno2014use} developed a new static technique, called Lobster, which calculated the similarity between code elements and source code from the stack traces. The experimental results on 14 projects showed that Lobster improved the effectiveness of Lucene-based bug localization in most cases.
Gong et al. \cite{gong2014locating} proposed a framework, which used the distance reweighting and test coverage reweighting techniques to locate post-release crashes based on stack traces. They conducted experiments on two versions of Firefox project and the results showed that their method could locate more than 63.9\% of crashing faults by examining 5\% of functions.
Wu et al. \cite{wu2018changelocator} proposed an automatic method, called ChangeLocator, which used features derived from crash reports and the historical fixed crashes to locate the crash-inducing changes. Their experiments on six versions of NetBeans project showed that ChangeLocator significantly outperformed the comparative methods.


\subsection{Feature Selection in Software Engineering}
The goal of feature selection methods is to choose an optimal feature subset to replace original ones by reserving the most representative features and removing the useless features for improving the performance of the machine learning model. Previous researchers introduced feature selection methods to relieve the issues in software engineering, such as software defect prediction and software effort estimation.

The objective of the software defect prediction task is to predict defective-prone software modules for software quality assurance. 
Liu et al. \cite{liu2014fecar} proposed a feature selection framework, called FECAR, which used feature clustering based on FF-Correlation measure and ranking relevant features based on FC-Relevance measure for software defect prediction task. Their results on Eclipse and NASA datasets showed the effectiveness of their method.
Chen et al. \cite{chen2014two} proposed a data preprocessing method which applied feature selection, threshold-based clustering, and random sampling techniques for defect prediction. Their results on NASA and Eclipse datasets showed that this method offered a solution for preprocessing cost-effective data.
Liu et al. \cite{liu2015fecs} proposed a FECS method, which used feature clustering and feature selection with three different search strategies. Their results on NASA and Eclipse datasets showed the effectiveness of this method for fault prediction with noises.
Ni et al. \cite{ni2019empirical} proposed a defect prediction method, called MOFES, which took both minimizing the number of selected features and maximizing the performance of models into account. Their experiments on RELINK and PROMISE datasets showed that MOFES provides a direction on collecting high-quality datasets for software defect prediction task. 
Cui et al. \cite{cui2019novel} proposed a novel feature selection method NFS, which applied the correlation-based feature subset selection and calculated the similarity of features to extract the useful features. They conducted experiments on 10 defect projects and the results showed the feasibility of the NFS method. 
Manjula et al. \cite{manjula2019deep} proposed a hybrid approach which combined the genetic algorithm and the deep neural network for feature learning and classification. Their results on PROMISE dataset showed that this method performed better than the comparative methods for predicting defects.
In addition, Xu et al. \cite{xu2016impact} and Ghotra et al. \cite{ghotra2017large} conducted empirical studies to investigate the impacts of feature selection on the performance of defect prediction models.

The objective of the software effort estimation is to estimate the number of resources consumed in the process of software development to assist the project management.
Azzeh et al. \cite{azzeh2008improving} proposed a fuzzy logic based feature subset selection method for improving the accuracy of software effort estimation model. They conducted experiments on ISBSG and Desharnais datasets and the results showed that their method performed significantly better than the hill climbing, forward subset selection, and backward subset selection. 
Oliveira et al. \cite{oliveira2010ga} proposed a genetic algorithm based method to select the optimal feature subset and optimize the model parameters at the same time for software effort estimation task. They conducted experiments on six datasets and the results showed that their method considerably reduced the number of original features and improved the performance of machine learning models.
Shahpar et al. \cite{shahpar2016improvement} employed a genetic algorithm to select useful features for software effort estimation task. Their experimental results on Desharnais, Maxwell, and CCOMO81 datasets showed that genetic algorithms are effective for improving the accuracy of the model.
Hosni et al. \cite{hosni2017investigating} investigated the impact of Correlation based Feature Selection (CFS) and RReliefF for the effort estimation accuracy of heterogeneous ensembles with four machine learning techniques. They conducted experiments on six datasets and the results showed that CFS ensembles
achieved better performance than RReliefF ensembles. 
Liu et al. \cite{liu2019feature} proposed a greedy feature selection method, called LFS, to guarantee the appropriateness of case based reasoning for software effort estimation task. They conducted experiments on six datasets and the results showed that the feature subset selected by LFS made effective estimation compared with a randomized baseline method.

Different from the above studies, we use the feature subset selection method to select an optimal feature subset from the crash data, expecting to build a high-quality training set.

\subsection{Deep Forest in Software Engineering}

Deep forest \cite{zhou2017deep} is a recently proposed forest-based ensemble method that consists of a multi-grained scanning and a cascade structure. Compared with deep neural networks, deep forest could obtain competitive performance on both large-scale and small-scale data with fewer hyper-parameters. Recently, two previous studies have used it to solve the problem of defect prediction in software engineering. 
Zhou et al. \cite{zhou2019improving} made the first attempt to build a deep forest based model for software defect prediction task by using z-score method to process the original features. The results on 25 projects showed that their method was more effective to identify defective software modules in terms of area under the receiver operating characteristic curve indicator. 
Zheng et al. \cite{zhengsoftware} employed an improved deep forest method based on data augmentation and autoencoder techniques for predicting software defects. They conducted experiments on Eclipse project and the results showed that their approach achieved better performance than original deep forest method. 
In this work, we are the first to introduce the deep forest model into the crashing fault residence prediction task.

\section{Method}\label{sec3}

\subsection{Overview}\label{3.1}

\begin{figure*}
	\includegraphics[width=0.75\paperwidth]{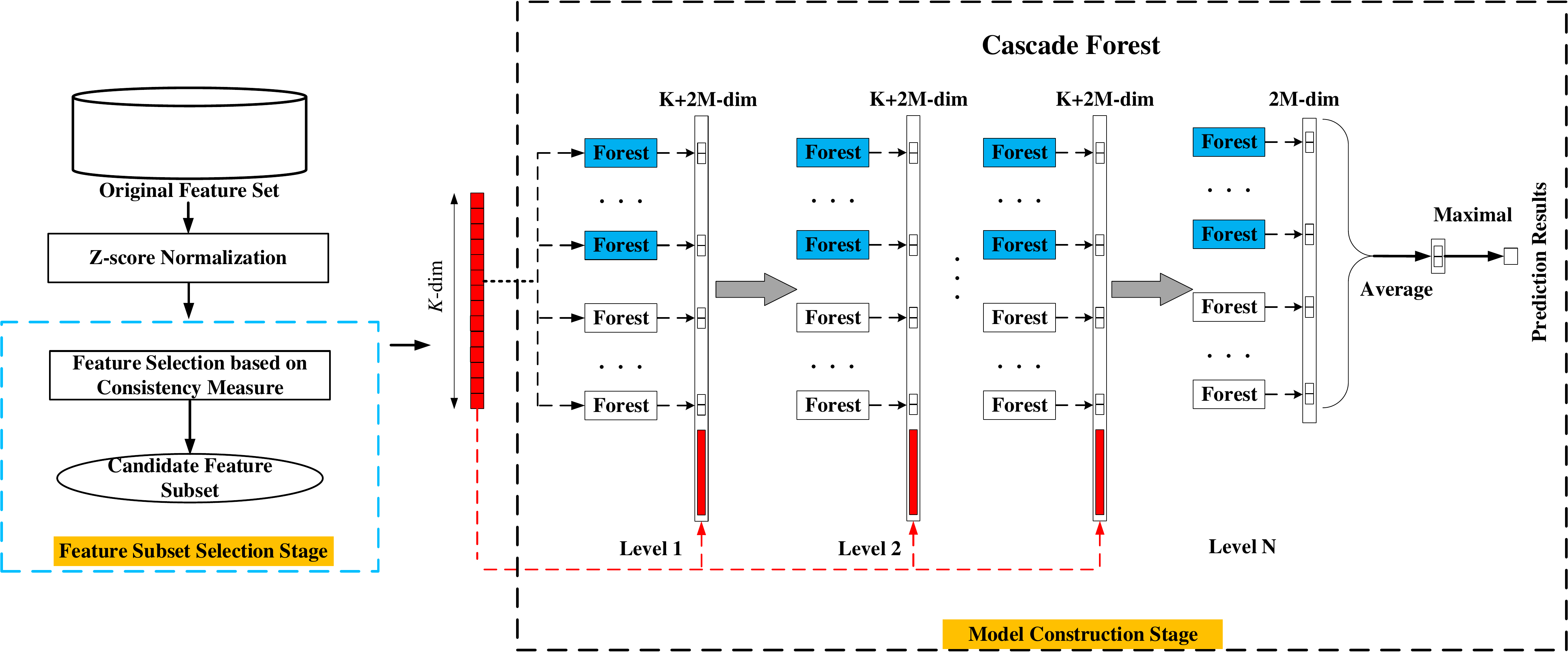}
	\centering
	\caption{An overview of our ConDF method}
	\label{frame_overview}
\end{figure*}


Figure \ref{frame_overview} demonstrates an overview of our proposed ConDF method, which consists of a feature reduction stage and a classification model construction stage. More specifically, in \textbf{step 1}, we first select the  important features from the original crash data as the candidate feature subset with consistency-based feature selection technique (short for Con). As a result, we can obtain a reduced feature set. In \textbf{step 2}, we apply a simplified deep forest technique (short for DF) to the reduced feature set for building the prediction model to identify whether the new crash instances reside in the stack trace or not. Before running our ConDF method, we first standardize the crash data with z-score method \cite{zhang2016cross} which normalizes the features with a mean value of zero and a variance of one. Below, we introduce the details of the used Con and the DF techniques, individually.

\subsection{Feature Reduction with Feature Subset Selection Method Con}
The objective of the feature subset selection or feature reduction stage is to single out the most useful features. For this purpose, in the first step, we apply the Con technique to select the feature subset. Con introduces the consistency measure as the criterion to evaluate the selected feature subset. The consistency is equal to 1 minus inconsistency which is defined as the proportion of the inconsistent samples in the total number of samples. The inconsistent instances are the ones that have the same feature value but different class labels \cite{liu1996probabilistic}\cite{dash2000consistency}. More specifically, Con calculates the consistency between a feature subset and class labels. It reserves a smallest set of features that has the same consistency as the whole feature set. In other words, the selected feature subset can distinguish classes as if with the whole feature set. Moreover, Con is a multivariate measure method which examines a subset of features at a time, and has low time complexity by using a hashing mechanism to calculate the inconsistency rate \cite{dash2000consistency}. Here, we take an example to show how to calculate the inconsistency. Assume that there are two different classes (i.e., $C_1$ and $C_2$) in the whole feature set $F$. For a given feature subset $S$ from $F$, there are $N_S$ instances with $s$ different values ($N_S=P_1+P_2+\cdots+P_s$), in which the instance numbers in $P_i (1\leq i \leq s)$ with label $C_1$ and $C_2$ are $N_1$ and $N_2$ ($N$=$N_1$+$N_2$) individually. If $N_1$ is the largest, the inconsistency rate of value $P_i$ is defined as $InP_i=\frac{N-N_1}{N_S}$. Then, the inconsistency can be defined as the sum of the inconsistency rate among all values.
After the processing of Con technique, we obtain an optimal feature subset as the candidate by removing the redundant and irrelevant features \cite{liu1996probabilistic}.

\subsection{Model Construction with Simplified DF Classifier}

DF \cite{zhou2017deep} is a novel decision tree ensemble approach inspired by the layer-by-layer structure in deep neural networks. It introduces different types of forests (such as random forests and completely-random tree forests) to learn the feature diversity. During the training process, the initial features of each instance are used many times to obtain the augmented features. DF has the potential for both representation learning and classification with a cascade structure. The ability of representation learning is enhanced by a multi-gained scanning technique. It is suitable to process high-dimensional features, such as image and sequence data. On the one hand, our crash data inherently does not have a large number of features. On the other hand, after the processing of feature selection, our crash data only contains low-dimensional features. Therefore, we apply the simplified version of deep forest only with the cascade structure to build the prediction model in this work. The cascade structure consists of many levels, and each level includes many forests. For an input vector, the forest produces the class distribution by averaging all estimation of the class to which it belongs. DF can obtain promising performance with relatively lower computational costs and fewer hyper-parameters compared with deep neural networks \cite{zhou2017deep}.

In the second step of our ConDF method, we input features selected by the previous step into the first level of the cascade structure. The output results from the previous level and the original features are taken as the input for the next level of the cascade structure. The number of cascade levels can be automatically determined by terminating the training procedure when there is no performance improvement \cite{zhou2017deep}. Here, we give an example to demonstrate the process of our ConDF. For the data of one project, we use Con method to select $K$ features from the training set where the selected features have the same consistency as the original feature set. Note that, the same features are reserved for the test set. The processed training set is input into DF for model training. More specifically, the reserved features are input in the first level of the cascade forest of the DF. After the processing of the first level, we can obtain $2M$ features. Then these features are concatenated with the initial $K$ features, (i.e., $K + 2M$ features as shown in Fig. \ref{frame_overview}), and input as a whole to the next level. After this iteration process is completed, ConDF predicts the label of each instance in the test set.

\section{Experimental Setup}\label{sec4}

\begin{table*}
	\centering
	\caption{The basic statistics of seven projects}
	\scalebox{0.92}[0.92]{
		\begin{tabular}{l l c c c c c c c}
			\hline
			Project & URL & Version & \# Mutants & \# Crashes & \# InTrace & \# OutTrace & \% Ratio \\
			\hline
			Apache Commons Codec (Codec) & https://github.com/apache/commons-codec & 1.10 & 2901 & 610 & 177 & 433 & 29.0\% \\
			\hline
			Apache Commons Collections (Colle) &  https://github.com/apache/commons-collections & 4.1 & 6650 & 1350 & 273 & 1077 & 20.2\% \\
			\hline
			Apache Commons IO (IO) & https://github.com/apache/commons-io & 2.5 & 3337 & 686 & 149 & 537 & 21.7\% \\
			\hline
			Jsoup & https://github.com/jhy/jsoup/ & 1.11.1 & 2657 & 601 & 120 & 481 & 20.0\% \\
			\hline
			JSqlParser (JSqlP) & http://github.com/JSQLParser/JSqlParser & 0.9.7 & 8757 & 647 & 61 & 586 & 9.4\% \\
			\hline
			Mango & https://github.com/jfaster/mango & 1.5.4 & 5149 & 733 & 53 & 680 & 7.2\% \\
			\hline
			Ormlite-Core (Ormli) & http://github.com/j256/ormlite-core & 5.1 & 3563 & 1303 & 326 & 977 & 25.0\% \\
			\hline
		\end{tabular}
	}
	\label{tab_project_info}
\end{table*}

\subsection{Dataset}

In this work, we employ a publicly available benchmark dataset collected by Gu et al. \cite{gu2019does} that includes 7 Java projects to evaluate the performance of our proposed ConDF method, including Apache Commons \textbf{Codec}, Apache Commons \textbf{Colle}ctions, Apache Commons \textbf{IO}, \textbf{Jsoup}, \textbf{JSqlP}arser, \textbf{Mango}, and \textbf{Ormli}te-Core. 
Codec contains the implementations of encoders and decoders for various formats.
Colle builds new interfaces and implementations with many powerful data structures to facilitate the development of most Java packages and applications.
IO provides many input and output related classes, such as streams, readers, writers, and files, to simplify the development of IO functionality.
Jsoup is a HTML parsing tool that contains the convenient interfaces to directly parse URL and other contents from real-world HTML.
JSqlP provides the simple solutions for many databases, which parses SQL statements and transforms them into a traversable hierarchy of Java classes.
Mango is a high-performance distributed based object relational mapping framework that simplifies the uses of relational database for object-oriented applications. 
Ormli contains the lightweight functionality for supporting Java database connectivity. 
The basic statistics of these projects is described in Table \ref{tab_project_info}, including the URL of GitHub repository (URL), the Version, the total number of mutants generated by projects (\# Mutants), the number of the kept mutants (\# Crashes), the number of crash instances inside the stack trace (\# InTrace) and outside the stack trace (\# OutTrace), and the ratio of \# InTrace to \# Crashes (\% Ratio). The main steps for collecting this benchmark dataset are described as follows: (1) \emph{Crash generation.} The PIT system\footnote{PIT: http://pitest.org/} is used to generate single-point mutations with 7 default mutators\footnote{Mutators: http://pitest.org/quickstart/mutators/\#INCREMENTS} to simulate the real-world crashes. Then, the mutations that maybe not produce crashes are filtered out using the following rules: the mutation passes all test cases and only the AssertionFailedError, ComparisonFailure, or test case is contained in exception stack traces. (2) \emph{Feature extraction.} In order to characterize the crashes, 89 features are extracted from the stack trace and the source code by Spoon\footnote{Spoon: http://spoon.gforge.inria.fr/} . These features belong to 5 groups, including features related to the \textbf{S}tack \textbf{T}race (ST), features extracted from the \textbf{T}op \textbf{C}lass/functions and the \textbf{B}ottom \textbf{C}lass/function in the frame (TC and BC), and features \textbf{N}ormalized by LOC (i.e., lines of codes) from \textbf{TC} and \textbf{BC} (NTC and NBC). Table \ref{tab_feature_info} briefly describes the definitions of these features. 
(3) \emph{Labeling crashes.} There are three major components in the frame of the stack trace, i.e., class name, function name, and line number. If the location of a crashing fault exactly matches the three components of any one frame in the stack trace, the crash instance is deemed to reside in the stack trace and labeled as `InTrace', otherwise, `OutTrace'. 

\begin{table}
	\centering
	\caption{The definitions of 89 features}
	\scalebox{0.75}[0.75]{
	\begin{tabular}{l l}
		\hline
		Feature & Description \\
		\hline
		\rowcolor[rgb]{ .749,  .749,  .749}\multicolumn{2}{l}{Feature Set ST - features related to the stack trace} \\
		ST01 & Type of the exception in the crash \\
		ST02 & Number of frames of the stack trace \\
		ST03 & Number of classes in the stack trace \\
		ST04 & Number of functions in the stack trace \\
		ST05 & Whether an overloaded function exists in the stack trace \\
		ST06 & Length of the name in the top class \\
		ST07 & Length of the name in the top function \\
		ST08 & Length of the name in the bottom class \\
		ST09 & Length of the name in the bottom function \\
		ST10 & Number of Java files in the project \\
		ST11 & Number of classes in the project \\
		\hline
		\rowcolor[rgb]{ .749,  .749,  .749}\multicolumn{2}{l}{Feature Set TC and BC - features extracted from the top (bottom) class/function in the frame} \\
		TC01(BC01) & Number of local variables \\
		TC02(BC02) & Number of fields \\
		TC03(BC03) & Number of except constructor functions \\
		TC04(BC04) & Number of imported packages \\
		TC05(BC05) & Whether the class is inherited from others \\
		TC06(BC06) & LOC of comments \\
		TC07(BC07) & LOC \\
		TC08(BC08) & Number of parameters \\
		TC09(BC09) & Number of local variables \\
		TC10(BC10) & Number of if-statements \\
		TC11(BC11) & Number of loops \\
		TC12(BC12) & Number of for statements \\
		TC13(BC13) & Number of for-each statements \\
		TC14(BC14) & Number of while statements \\
		TC15(BC15) & Number of do-while statements \\
		TC16(BC16) & Number of try blocks \\
		TC17(BC17) & Number of catch blocks \\
		TC18(BC18) & Number of finally blocks \\
		TC19(BC19) & Number of assignment statements \\
		TC20(BC20) & Number of function calls \\
		TC21(BC21) & Number of return statements \\
		TC22(BC22) & Number of unary operators \\
		TC23(BC23) & Number of binary operators \\
		\hline
		\rowcolor[rgb]{ .749,  .749,  .749}\multicolumn{2}{l}{Feature Set NTC and NBC - features normalized by LOC from Feature Set TC and BC} \\
		NTC01(NBC01) & TC08/TC07(BC08/CBC07) \\
		NTC02(NBC02) & TC09/TC07(BC09/CBC07) \\
		...   \\
		NTC16(NBC16) & TC23/TC07(BC23/CBC07) \\
		\hline
	\end{tabular}
	}
	\label{tab_feature_info}
\end{table}

\subsection{Performance Indicators}

As the goal of this work is to predict a crash instance as `InTrace' or `OutTrace', it is a binary classification problem. In this work, we employ F-measure and \textbf{M}atthews \textbf{C}orrelation \textbf{C}oefficient (MCC) as indicators to evaluate the performance of our ConDF framework for identifying the crashing fault residence. We first introduce four basic terms widely used in the binary classification scenarios. \textbf{T}rue \textbf{P}ositive (TP) means the number of crash instances labeled as `InTrace' that are predicted as `InTrace'. \textbf{F}alse \textbf{P}ositive (FP) means the number of instances labeled as `OutTrace' that are predicted as `InTrace'. \textbf{F}alse \textbf{N}egative (FN) means the number of instances labeled as `InTrace' that are predicted as `OutTrace', and \textbf{T}rue \textbf{N}egative (TN) means the number of instances labeled as `OutTrace' that are predicted as `OutTrace'. The F-measure for crash instances with label `InTrace', short for $\rm{F_{InTrace}}$, is defined as the following formula.
\begin{equation}
	\rm{F_{InTrace}} = \frac{2 \times  \rm{Precision} \times \rm{Recall}}{\rm{Precision} + \rm{Recall}}
\end{equation}
where Precision=$\rm{\frac{TP}{TP+FP}}$ and Recall=$\rm{\frac{TP}{TP+FN}}$.
F-measure is a trade-off between Precision and Recall. In terms of the F-measure for crashing faults with the label `OutTrace', we can obtain the similar expressions TP$\rm{_O}$ (=TN), FP$\rm{_O}$ (=FN), FN$\rm{_O}$ (=FP), and TN$\rm{_O}$ (=TP). Then, the F-measure for crashing faults with label `OutTrace', short for $\rm{F_{OutTrace}}$, is defined as the following formula.
\begin{equation}
	\rm{F_{OutTrace}} = \frac{2 \times  \rm{Precision_O} \times \rm{Recall_O}}{\rm{Precision_O} + \rm{Recall_O}}
\end{equation}
where Precision$\rm{_O}$=$\rm{\frac{TP_O}{TP_O+FP_O}}$ and Recall$\rm{_O}$=$\rm{\frac{TP_O}{TP_O+FN_O}}$.

MCC is a correlation coefficient considering TP, FP, FN, and TN, which is always used to measure the performance of binary classification. MCC for crashing faults with label `InTrace' is defined as the following formula.
\begin{equation}
	\small
	\rm{MCC} = \frac{\rm{TP} \times \rm{TN} - \rm{FP} \times \rm{FN}}{\sqrt{(\rm{TP} + \rm{FP})(\rm{TP} + \rm{FN})(\rm{TN} + \rm{FP})(\rm{TN} + \rm{FN})}}
\end{equation}

According to the correlation between TN, TP, FN, FP and TN$\rm{_O}$, TP$\rm{_O}$, FN$\rm{_O}$, FP$\rm{_O}$, MCC for crash instances with label `InTrace' is the same as MCC for crash instances with label `OutTrace'. 

$\rm{F_{InTrace}}$ and $\rm{F_{OutTrace}}$ range from 0 to 1. MCC ranges from -1 to 1. The larger indicator value indicates better performance. MCC = -1 means the worst prediction and MCC = 1 means the perfect prediction. MCC = 0 denotes that the performance is equal to random prediction. 

\subsection{Data Partition}
In this work, we employ the stratified sampling technique to generate the training set and test set. More specifically, for each project, the training set contains half instances of the crash data with label `InTrace' and `OutTrace', and the test set contains the remaining ones. The stratified sampling technique makes the proportions of crash data with label `InTrace' and `OutTrace' in the training set and test set are the same as that in the original one. To alleviate the bias of random partition, we repeat this process 50 times and report the corresponding average values and standard deviations with each indicator. 

\subsection{Parameter Settings}
In this work, we need to specify two parameters, 
the number of forests $M$ in each level of the cascade structure and the number of trees in each forest. 
For the first parameter, we construct the deep forest in which each level of the cascade consists of four random forests and four completely-random tree forests, i.e., we set $M$ as 8. In addition, we set each forest with 500 trees. The settings for  these two parameters follow the work in \cite{zhou2017deep}.  Note that, the number of features retained in the first stage is determined by Con technique automatically. 

\subsection{Statistic Test}
\begin{figure}
	\includegraphics[width=\linewidth]{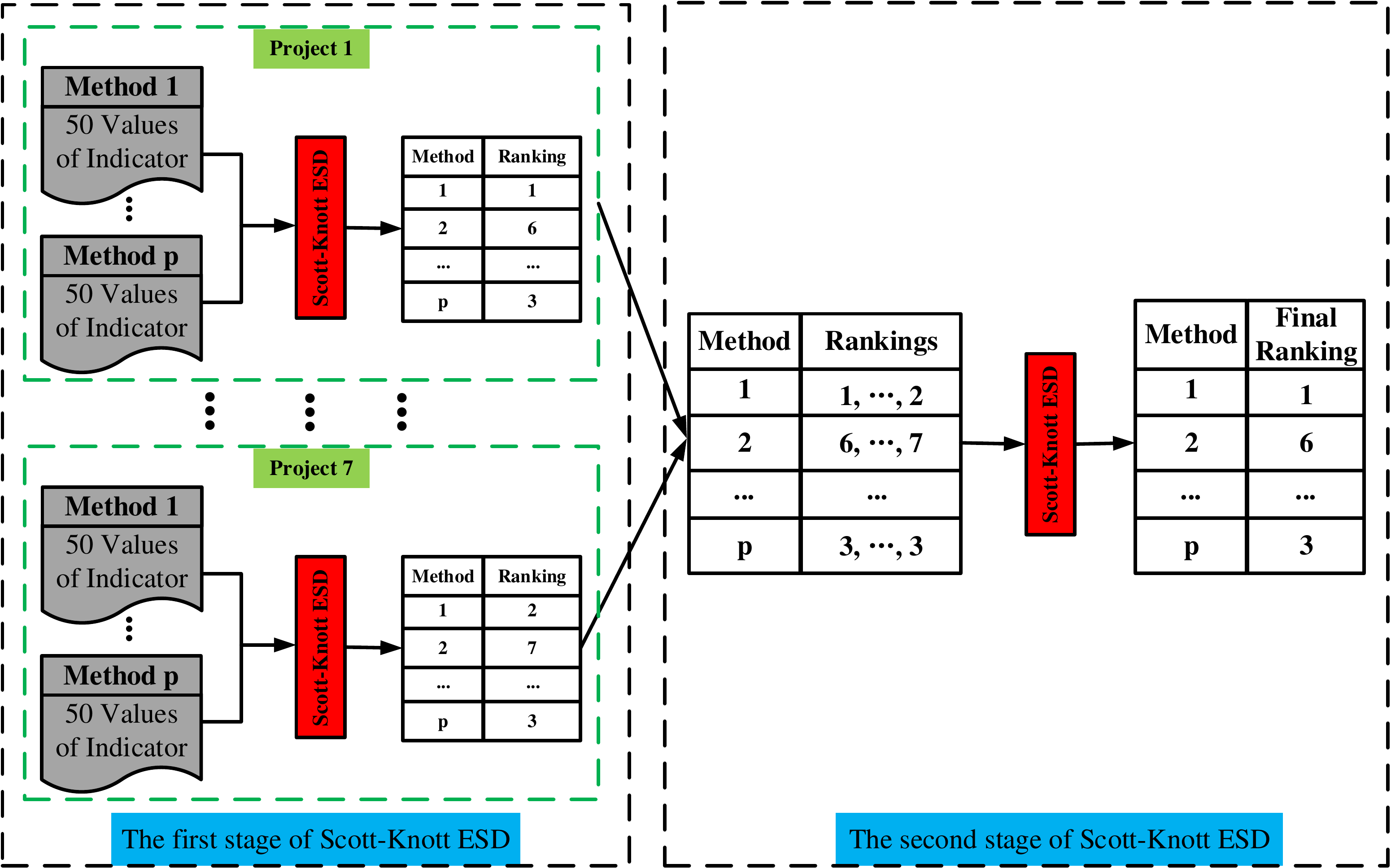}
	\caption{The process of Scott-Knott ESD test}
	\label{skesd}
\end{figure}
In this work, we apply a state-of-the-art statistical test method, called \textbf{Scott-Knott E}ffect \textbf{S}ize \textbf{D}ifference (Scott-Knott ESD) \cite{tantithamthavorn2016empirical}, to analyze the significant differences between our ConDF method and the comparative methods. Different from the original Scott-Knott test, Scott-Knott ESD applies the log-transforming to correct the non-normal distribution of inputs and merges the groups with a negligible effect size of differences. Fig. \ref{skesd} illustrates the analysis process of this test, which contains two stages. In the first stage, Scott-Knott ESD takes the 50 random indicator values of each method on each project as input. The output is the ranking value of each method on each project. In the second stage, this test method takes the output from the first stage as input and the output is the final ranking of each method across all projects. The low ranking means that the corresponding method achieves better indicator values. The methods with distinct colors mean that they are significantly different with a significance level $\alpha$ = 0.05.

\section{Experimental Results}\label{sec5}
\subsection{RQ1: Is the prediction performance of our proposed ConDF method better than that of ensemble based methods?}

\textbf{Motivation}:
As the DF technique used in the second step of our proposed framework is a cascade forest ensemble method, thus our ConDF method is a novel kind of the ensemble based method. Ensemble learning improves the performance by integrating multiple weak classifiers as a stronger classifier. 
This question is designed to investigate whether our ConDF framework with DF is better than some ensemble based methods with the Con method in the performance improvement for predicting crashing fault residence.

\textbf{Methods}:
To answer this question, we choose five ensemble based methods, including \textbf{Bag}ging (Bag), \textbf{B}alanced \textbf{Bag}ging (BBag), \textbf{Ada}ptive \textbf{B}oosting (AdaB), \textbf{R}andom \textbf{U}nder-\textbf{S}ampling with Ada\textbf{B} (RUSB), and \textbf{B}alanced \textbf{R}andom \textbf{F}orest (BRF), and combine these methods with Con for comparison, short for ConBag, ConBBag, ConAdaB, ConRUSB, and ConBRF, respectively. In addition, we add the state-of-the-art method CraTer \cite{gu2019does} for the crash fault residence prediction task as the basic method for comparison.

\begin{table}[htbp]
	\centering
	\caption{The Average $\rm{F_{InTrace}}$ of ConDF, other ensemble based methods with Con, and the state-of-the-art method}
	\scalebox{0.65}[0.65]{
    \begin{tabular}{c|ccccccc}
    \toprule
    Project & ConBag   & ConBBag & ConAdaB  & ConRUSB  & ConBRF & CraTer & ConDF \\
    \midrule
    Codec & 0.668(0.07) & 0.661(0.06) & 0.625(0.10) & 0.565(0.09) & \textbf{0.703}(0.06) & 0.612(0.06) & 0.678(0.09) \\
    Colle & 0.627(0.04) & 0.596(0.03) & 0.724(0.04) & 0.639(0.06) & 0.713(0.03) & 0.520(0.09) & \textbf{0.774}(0.04) \\
    IO    & 0.714(0.04) & 0.666(0.04) & 0.743(0.05) & 0.691(0.07) & 0.727(0.04) & 0.651(0.10) & \textbf{0.763}(0.05) \\
    Jsoup & 0.521(0.09) & 0.514(0.07) & 0.503(0.12) & 0.438(0.09) & \textbf{0.538}(0.06) & 0.473(0.07) & 0.521(0.13) \\
    JSqlP & \textbf{0.720}(0.06) & 0.575(0.10) & 0.672(0.13) & 0.581(0.17) & 0.528(0.08) & 0.496(0.14) & 0.684(0.19) \\
    Mango & 0.671(0.15) & 0.498(0.21) & 0.675(0.18) & 0.584(0.17) & 0.510(0.18) & 0.418(0.19) & \textbf{0.761}(0.10) \\
    Ormli & 0.701(0.04) & 0.683(0.03) & 0.853(0.08) & 0.779(0.10) & 0.829(0.06) & 0.710(0.10) & \textbf{0.870}(0.03) \\
    \midrule
    Average & 0.660(0.06) & 0.599(0.07) & 0.685(0.10) & 0.611(0.10) & 0.650(0.12) & 0.554(0.10) & \textbf{0.722}(0.10) \\
    \bottomrule
    \end{tabular}%
	}
	\label{RQ1-F}%
\end{table}%

\begin{table}[htbp]
	\centering
	\caption{The Average $\rm{F_{OutTrace}}$ of ConDF, other ensemble based methods with Con, and the state-of-the-art method}
	\scalebox{0.65}[0.65]{
    \begin{tabular}{c|ccccccc}
    \toprule
    Project & ConBag   & ConBBag & ConAdaB  & ConRUSB  & ConBRF & CraTer & ConDF \\
    \midrule
    Codec & 0.871(0.02) & 0.839(0.03) & 0.869(0.03) & 0.841(0.03) & 0.853(0.03) & 0.768(0.07) & \textbf{0.879}(0.03) \\
    Colle & 0.917(0.01) & 0.862(0.02) & 0.939(0.01) & 0.913(0.01) & 0.914(0.02) & 0.738(0.12) & \textbf{0.950}(0.01) \\
    IO    & 0.925(0.01) & 0.880(0.02) & 0.936(0.01) & 0.922(0.01) & 0.909(0.02) & 0.852(0.12) & \textbf{0.940}(0.01) \\
    Jsoup & 0.899(0.02) & 0.841(0.04) & 0.895(0.04) & 0.868(0.03) & 0.833(0.04) & 0.790(0.10) & \textbf{0.909}(0.03) \\
    JSqlP & \textbf{0.975}(0.01) & 0.936(0.03) & 0.969(0.01) & 0.959(0.01) & 0.917(0.04) & 0.881(0.13) & \textbf{0.975}(0.01) \\
    Mango & 0.979(0.01) & 0.913(0.07) & 0.975(0.02) & 0.959(0.03) & 0.923(0.05) & 0.835(0.18) & \textbf{0.982}(0.02) \\
    Ormli & 0.907(0.01) & 0.860(0.02) & 0.954(0.02) & 0.934(0.02) & 0.937(0.02) & 0.85(0.09) & \textbf{0.960}(0.01) \\
    \midrule
    Average & 0.925(0.04) & 0.876(0.03) & 0.934(0.04) & 0.914(0.04) & 0.898(0.04) & 0.816(0.05) & \textbf{0.942}(0.03) \\
    \bottomrule
    \end{tabular}%
	}
	\label{RQ1-FN}%
\end{table}%

\begin{table}[htbp]
	\centering
	\caption{The Average MCC of ConDF, other ensemble based methods with Con, and the state-of-the-art method.}
	\scalebox{0.65}[0.65]{
    \begin{tabular}{c|ccccccc}
    \toprule
    Project & Bag   & BalBag & AdaB  & RUSB  & BalRF & CraTer & ConDF \\
    \midrule
    Codec & 0.544(0.08) & 0.515(0.07) & 0.512(0.11) & 0.419(0.10) & \textbf{0.572(0.08)} & 0.427(0.10) & 0.567(0.10) \\
    Colle & 0.552(0.05) & 0.483(0.04) & 0.672(0.05) & 0.560(0.06) & 0.638(0.04) & 0.384(0.13) & \textbf{0.735}(0.04) \\
    IO    & 0.642(0.05) & 0.568(0.06) & 0.687(0.06) & 0.621(0.08) & 0.648(0.05) & 0.545(0.14) & \textbf{0.712}(0.05) \\
    Jsoup & 0.433(0.10) & 0.377(0.10) & 0.424(0.13) & 0.319(0.11) & 0.410(0.08) & 0.315(0.11) & \textbf{0.472}(0.14) \\
    JSqlP & \textbf{0.711}(0.06) & 0.541(0.10) & 0.653(0.13) & 0.550(0.18) & 0.498(0.08) & 0.447(0.17) & 0.685(0.18) \\
    Mango & 0.669(0.15) & 0.482(0.21) & 0.662(0.19) & 0.562(0.18) & 0.496(0.18) & 0.391(0.20) & \textbf{0.761}(0.10) \\
    Ormli & 0.611(0.05) & 0.570(0.04) & 0.810(0.09) & 0.721(0.10) & 0.772(0.08) & 0.610(0.13) & \textbf{0.833}(0.04) \\
    \midrule
    Average & 0.595(0.09) & 0.505(0.06) & 0.631(0.12) & 0.536(0.12) & 0.576(0.11) & 0.446(0.10) & \textbf{0.681}(0.11) \\
    \bottomrule
    \end{tabular}%
	}
	\label{RQ1-MCC}%
\end{table}%

\begin{figure}[htbp]
	\centering
	\subfigure[$\rm{F_{InTrace}}$]{
		\label{fig:RQ1-F}
		\begin{minipage}{0.9\linewidth}
			\centering
			\includegraphics[width=1\linewidth]{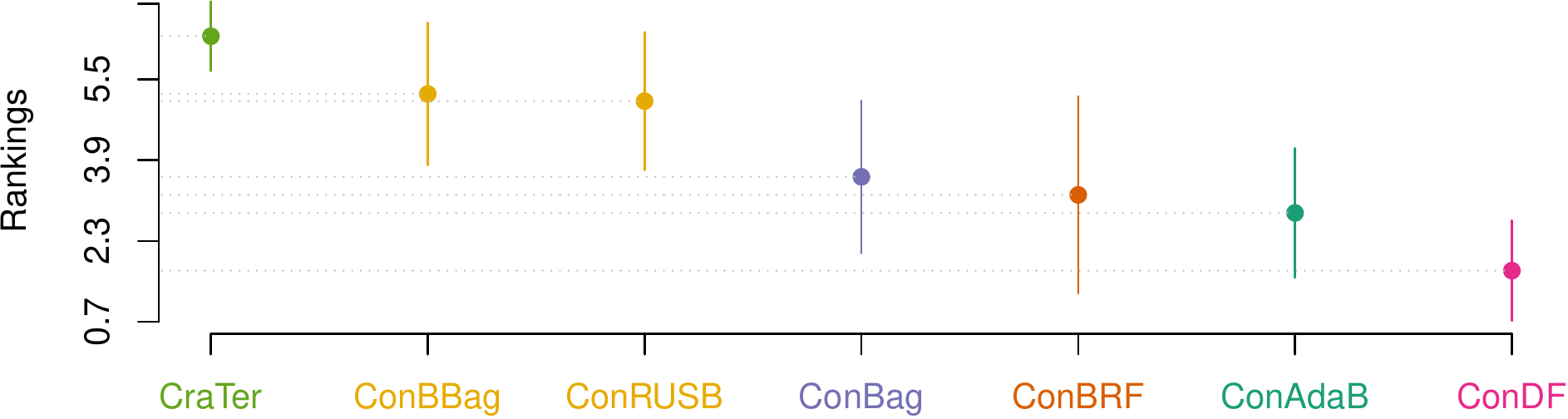}
		\end{minipage}
	}
	\subfigure[$\rm{F_{OutTrace}}$]{
		\label{fig:RQ1-FN}
		\begin{minipage}{0.9\linewidth}
			\centering
			\includegraphics[width=1\linewidth]{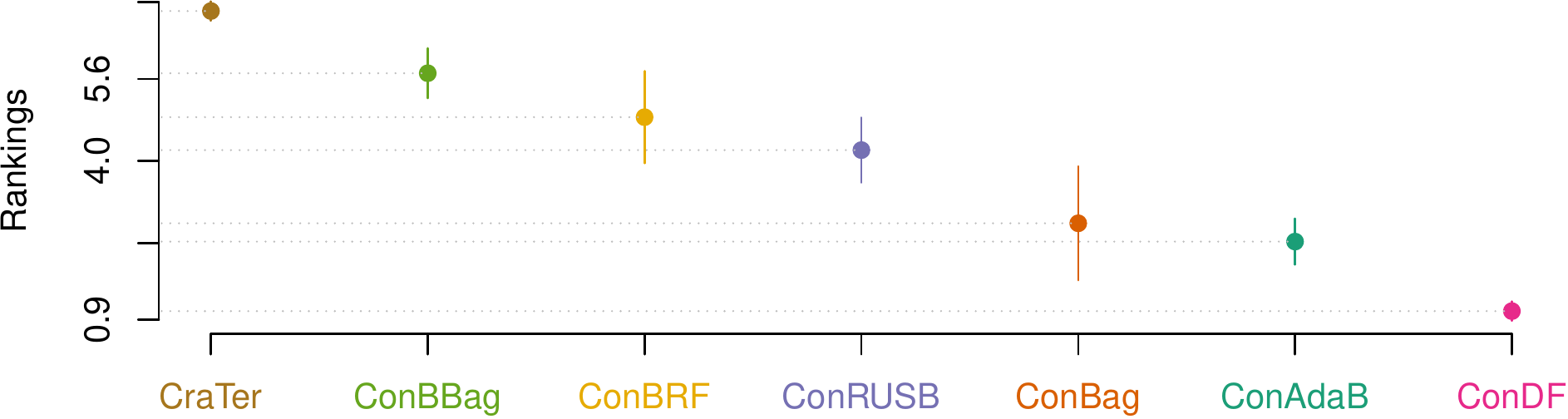}
		\end{minipage}
	}	
	\subfigure[MCC]{
		\label{fig:RQ1-MCC}
		\begin{minipage}{0.9\linewidth}
			\centering
			\includegraphics[width=1\linewidth]{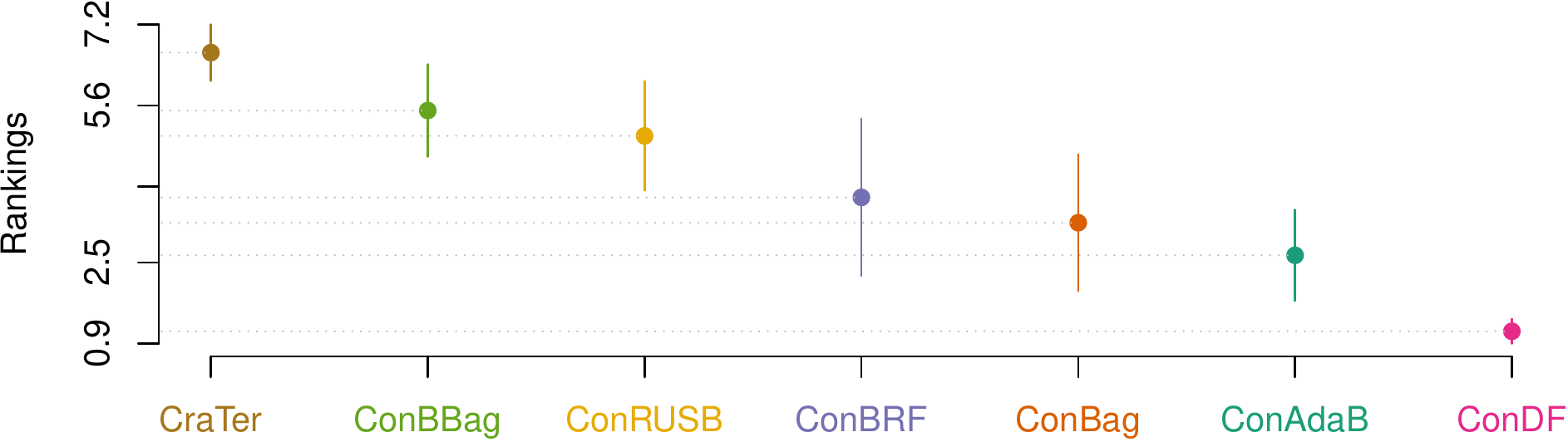}
		\end{minipage}
	}
	\caption{Scott-Knott ESD test for ConDF, other ensemble based methods with Con, and the state-of-the-art method.}
	\label{fig:RQ1-CK-SKESD}
\end{figure}

\textbf{Results}:
Table \ref{RQ1-F}, \ref{RQ1-FN}, and \ref{RQ1-MCC} report the results of the indicator values for our ConDF method and the 6 comparative methods in terms of $\rm{F_{InTrace}}$, $\rm{F_{OutTrace}}$, and MCC, individually, including the average values of the 50 random runs and the corresponding standard deviations in the brackets. 
In terms of $\rm{F_{InTrace}}$, ConDF obtains better performance on 4 out of 7 projects compared with the 6 baseline methods. The average $\rm{F_{InTrace}}$ value by our ConDF method over all projects achieves improvements by 9.4\%, 20.5\%, 5.4\%, 18.2\%, 11.1\%, and 30.3\% compared with ConBag, ConBBag, ConAdaB, ConRUSB, ConBRF, and CraTer, individually. 
In terms of $\rm{F_{OutTrace}}$, ConDF obtains better performance on 6 out of 7 projects compared with the 6 baseline methods. The average $\rm{F_{OutTrace}}$ value by our ConDF method over all projects achieves improvements by 1.8\%, 7.5\%, 0.9\%, 3.1\%, 4.9\%, and 15.4\% compared with the 6 baseline methods, individually. 
In terms of MCC, ConDF obtains better performance on 5 out of 7 projects compared with the 6 baseline methods. The average MCC value by our ConDF method over all projects achieves improvements by 14.5\%, 34.9\%, 7.9\%, 27.1\%, 18.2\%, and 52.7\% compared with the 6 baseline methods, individually. 
Overall, our proposed ConDF method obtains the best average value and achieves average improvements by 15.8\%, 5.6\%, and 25.9\% in terms of $\rm{F_{InTrace}}$, $\rm{F_{OutTrace}}$, and MCC, individually.

Fig. \ref{fig:RQ1-CK-SKESD} visualizes the corresponding statistical test results for our ConDF method and the 6 comparative methods in terms of all three indicators. This figure illustrates that our ConDF method always ranks the first and has significant differences compared with the baseline methods in terms of all indicators. 

\textbf{Answer}:
Our proposed ConDF method performs significantly better than the comparative ensemble based methods with feature selection method Con and the state-of-the-art method CraTer for predicting crashing fault residence in terms of all three indicators. 

\begin{table}[htbp]
	\centering
	\caption{The Average $\rm{F_{InTrace}}$ of ConDF and Con combining other classifiers}
	\scalebox{0.73}[0.73]{
		\begin{tabular}{c|cccccc}
			\toprule
			Project & ConDT    & ConSVM   & ConLR    & ConRF    & ConNN    & ConDF \\
			\midrule
			Codec & \textbf{0.684}(0.07) & 0.414(0.07) & 0.549(0.05) & 0.143(0.08) & 0.601(0.04) & 0.678(0.09) \\
			Colle & 0.744(0.06) & 0.455(0.05) & 0.565(0.03) & 0.110(0.08) & 0.543(0.03) & \textbf{0.774}(0.04) \\
			IO    & 0.721(0.06) & 0.718(0.04) & 0.713(0.03) & 0.448(0.07) & 0.691(0.04) & \textbf{0.763}(0.05) \\
			Jsoup & 0.501(0.08) & 0.271(0.09) & 0.445(0.06) & 0.101(0.09) & 0.403(0.06) & \textbf{0.521}(0.13) \\
			JSqlP & 0.657(0.14) & 0.701(0.06) & \textbf{0.745}(0.05) & 0.600(0.26) & 0.714(0.07) & 0.684(0.19) \\
			Mango & 0.570(0.10) & 0.430(0.11) & 0.635(0.08) & 0.167(0.14) & 0.570(0.08) & \textbf{0.761}(0.10) \\
			Ormli & 0.821(0.04) & 0.657(0.03) & 0.651(0.03) & 0.236(0.12) & 0.651(0.03) & \textbf{0.870}(0.03) \\
			\midrule
			Average & 0.671(0.10) & 0.521(0.16) & 0.615(0.10) & 0.258(0.18) & 0.596(0.10) & \textbf{0.722}(0.10) \\
			\bottomrule
		\end{tabular}%
	}
	\label{RQ2-F}%
\end{table}%

\begin{table}[htbp]
	\centering
	\caption{The Average $\rm{F_{OutTrace}}$ of ConDF and Con combining other classifiers}
	\scalebox{0.73}[0.73]{
		\begin{tabular}{c|cccccc}
			\toprule
			Project & ConDT    & ConSVM   & ConLR    & ConRF    & ConNN    & ConDF \\
			\midrule
			Codec & 0.870(0.03) & 0.834(0.01) & 0.826(0.02) & 0.831(0.01) & 0.834(0.02) & \textbf{0.879}(0.03) \\
			Colle & 0.937(0.01) & 0.906(0.01) & 0.908(0.01) & 0.889(0.00) & 0.893(0.01) & \textbf{0.950}(0.01) \\
			IO    & 0.925(0.01) & 0.931(0.01) & 0.924(0.01) & 0.902(0.01) & 0.912(0.01) & \textbf{0.940}(0.01) \\
			Jsoup & 0.875(0.03) & 0.893(0.01) & 0.881(0.01) & 0.890(0.00) & 0.877(0.02) & \textbf{0.909}(0.03) \\
			JSqlP & 0.947(0.07) & \textbf{0.976}(0.00) & 0.975(0.01) & 0.972(0.01) & \textbf{0.976}(0.01) & 0.975(0.01) \\
			Mango & 0.968(0.01) & 0.970(0.00) & 0.974(0.01) & 0.965(0.00) & 0.973(0.01) & \textbf{0.982}(0.02) \\
			Ormli & 0.941(0.01) & 0.906(0.01) & 0.897(0.01) & 0.868(0.01) & 0.884(0.01) & \textbf{0.960}(0.01) \\
			\midrule
			Average & 0.923(0.03) & 0.917(0.05) & 0.912(0.05) & 0.902(0.05) & 0.907(0.05) & \textbf{0.942}(0.03) \\
			\bottomrule
		\end{tabular}%
	}
	\label{RQ2-FN}%
\end{table}%

\begin{table}[htbp]
	\centering
	\caption{The Average MCC of ConDF and Con combining other classifiers}
	\scalebox{0.73}[0.73]{
		\begin{tabular}{c|cccccc}
			\toprule
			Project & ConDT    & ConSVM   & ConLR    & ConRF    & ConNN    & ConDF \\
			\midrule
			Codec & 0.561(0.10) & 0.297(0.05) & 0.378(0.06) & 0.136(0.06) & 0.440(0.05) & \textbf{0.567}(0.10) \\
			Colle & 0.684(0.07) & 0.421(0.05) & 0.485(0.04) & 0.135(0.08) & 0.440(0.04) & \textbf{0.735}(0.04) \\
			IO    & 0.650(0.07) & 0.659(0.04) & 0.639(0.04) & 0.431(0.06) & 0.605(0.05) & \textbf{0.712}(0.05) \\
			Jsoup & 0.383(0.09) & 0.258(0.07) & 0.334(0.06) & 0.121(0.10) & 0.292(0.08) & \textbf{0.472}(0.14) \\
			JSqlP & 0.624(0.16) & 0.715(0.05) & \textbf{0.727}(0.06) & 0.615(0.25) & 0.713(0.06) & 0.685(0.18) \\
			Mango & 0.546(0.11) & 0.468(0.18) & 0.618(0.09) & 0.232(0.18) & 0.571(0.09) & \textbf{0.761}(0.10) \\
			Ormli & 0.764(0.05) & 0.580(0.04) & 0.554(0.03) & 0.258(0.09) & 0.536(0.04) & \textbf{0.833}(0.04) \\
			\midrule
			Average & 0.602(0.11) & 0.485(0.16) & 0.534(0.13) & 0.275(0.17) & 0.514(0.13) & \textbf{0.681}(0.11) \\
			\bottomrule
		\end{tabular}%
		\label{RQ2-MCC}%
	}
\end{table}%

\begin{figure}[htbp]
	\centering
	\subfigure[$\rm{F_{InTrace}}$]{
		\label{fig:RQ2-F}
		\begin{minipage}{0.9\linewidth}
			\centering
			\includegraphics[width=1\linewidth]{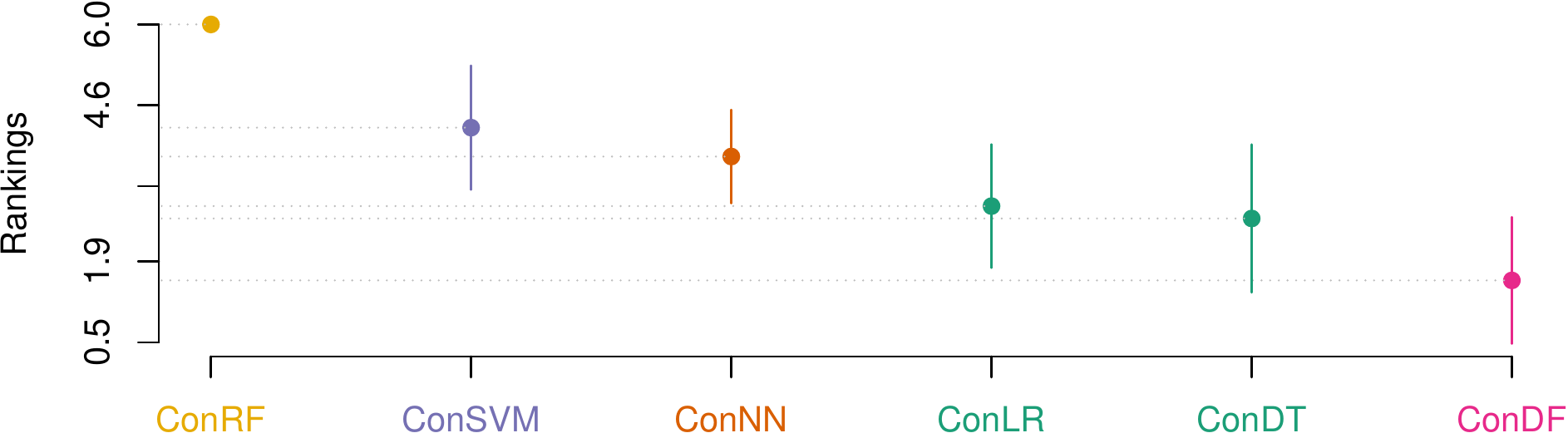}
		\end{minipage}
	}
	\subfigure[$\rm{F_{OutTrace}}$]{
		\label{fig:RQ2-FN}
		\begin{minipage}{0.9\linewidth}
			\centering
			\includegraphics[width=1\linewidth]{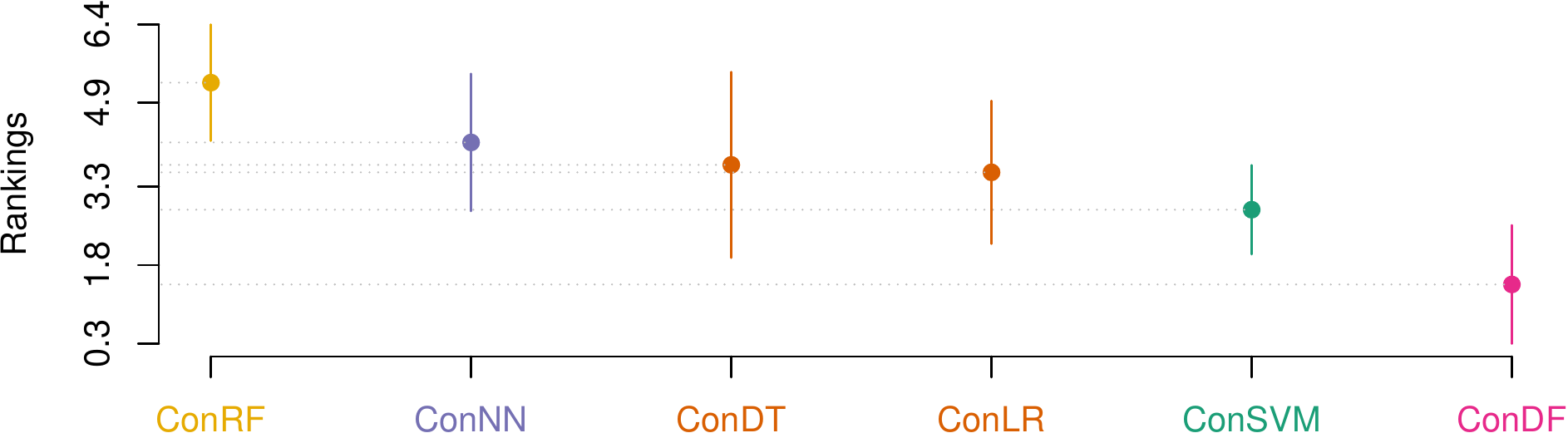}
		\end{minipage}
	}	
	\subfigure[MCC]{
		\label{fig:RQ2-MCC}
		\begin{minipage}{0.9\linewidth}
			\centering
			\includegraphics[width=1\linewidth]{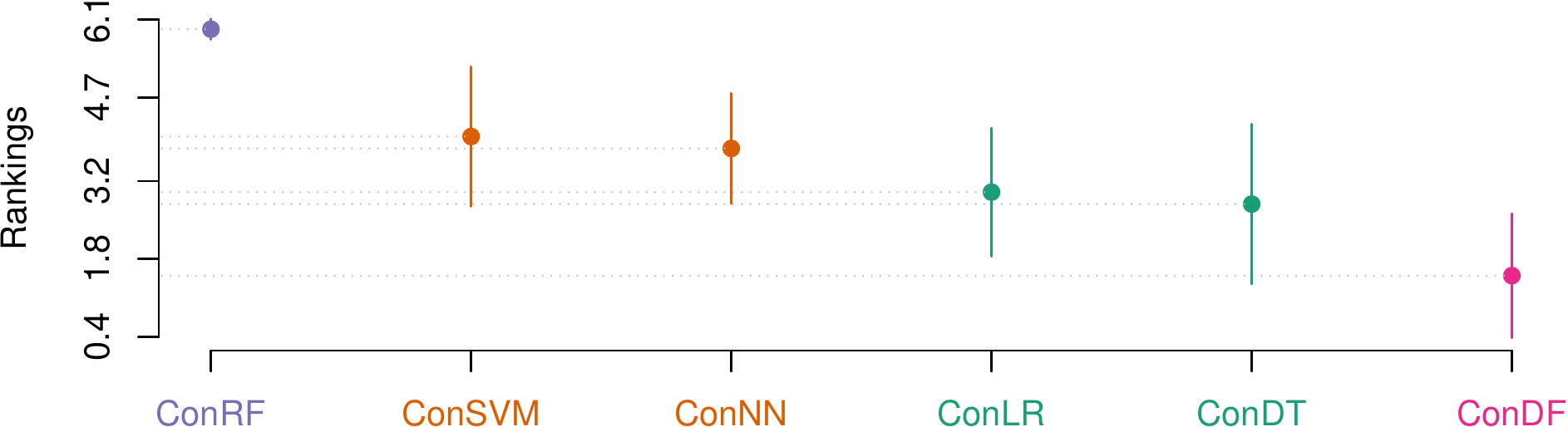}
		\end{minipage}
	}

	\caption{Scott-Knott ESD test for ConDF and Con combining other classifiers.}
	\label{fig:RQ2-CK-SKESD}
\end{figure}

\begin{table}[htbp]
	\centering
	\caption{The Average $\rm{F_{InTrace}}$ of ConDF and other feature selection methods using DF classifier}
	\scalebox{0.65}[0.65]{
	    \begin{tabular}{c|ccccccc}
	    \toprule
	    Project & CSDF  & IGDF  & GRDF  & ReFDF & CFSDF & DF    & ConDF \\
	    \midrule
	    Codec & 0.647(0.10) & 0.664(0.08) & 0.669(0.09) & 0.573(0.11) & 0.656(0.09) & 0.615(0.08) & \textbf{0.678}(0.09) \\
	    Colle & 0.766(0.04) & 0.760(0.04) & 0.768(0.07) & 0.715(0.05) & \textbf{0.776}(0.03) & 0.729(0.03) & 0.774(0.04) \\
	    IO    & 0.715(0.07) & 0.722(0.07) & 0.716(0.06) & 0.714(0.09) & 0.710(0.09) & 0.743(0.05) & \textbf{0.763}(0.05) \\
	    Jsoup & 0.471(0.12) & 0.468(0.13) & 0.474(0.12) & \textbf{0.539}(0.13) & 0.516(0.13) & 0.437(0.11) & 0.521(0.13) \\
	    JSqlP & 0.659(0.21) & 0.555(0.32) & 0.604(0.27) & 0.678(0.21) & 0.581(0.28) & 0.718(0.15) & \textbf{0.684}(0.19) \\
	    Mango & 0.681(0.09) & 0.662(0.13) & 0.688(0.10) & 0.345(0.16) & \textbf{0.767}(0.06) & 0.608(0.12) & 0.761(0.10) \\
	    Ormli & 0.781(0.06) & 0.779(0.06) & 0.810(0.05) & 0.790(0.06) & 0.838(0.05) & 0.737(0.05) & \textbf{0.870}(0.03) \\
	    \midrule
	    Average & 0.674(0.10) & 0.659(0.10) & 0.676(0.10) & 0.622(0.14) & 0.692(0.11) & 0.655(0.10) & \textbf{0.722}(0.10) \\
	    \bottomrule
	    \end{tabular}%
	}
	\label{RQ3-F}%
\end{table}%

\begin{table}[htbp]
	\centering
	\caption{The Average $\rm{F_{OutTrace}}$ of ConDF and other feature selection methods using DF classifier}
	\scalebox{0.65}[0.65]{
	    \begin{tabular}{c|ccccccc}
	    \toprule
	    Project & CSDF  & IGDF  & GRDF  & ReFDF & CFSDF & DF    & ConDF \\
	    \midrule
	    Codec & 0.865(0.03) & 0.868(0.03) & 0.873(0.03) & 0.845(0.04) & 0.874(0.03) & 0.852(0.03) & \textbf{0.879}(0.03) \\
	    Colle & 0.948(0.01) & 0.946(0.01) & 0.949(0.01) & 0.942(0.01) & \textbf{0.952}(0.01) & 0.940(0.01) & 0.950(0.01) \\
	    IO    & 0.930(0.01) & 0.932(0.01) & 0.926(0.02) & 0.930(0.01) & 0.929(0.02) & 0.936(0.01) & \textbf{0.940}(0.01) \\
	    Jsoup & 0.905(0.01) & 0.905(0.02) & 0.908(0.01) & \textbf{0.913}(0.01) & 0.911(0.03) & 0.901(0.01) & 0.909(0.03) \\
	    JSqlP & 0.974(0.01) & 0.970(0.01) & 0.972(0.01) & 0.976(0.01) & 0.971(0.01) & \textbf{0.977}(0.01) & 0.975(0.01) \\
	    Mango & 0.979(0.01) & 0.979(0.01) & 0.980(0.01) & 0.964(0.01) & \textbf{0.985}(0.00) & 0.976(0.01) & 0.982(0.02) \\
	    Ormli & 0.937(0.01) & 0.935(0.02) & 0.941(0.02) & 0.938(0.02) & 0.950(0.01) & 0.926(0.01) & \textbf{0.960}(0.01) \\
	    \midrule
	    Average & 0.934(0.04) & 0.934(0.04) & 0.936(0.03) & 0.930(0.04) & 0.939(0.04) & 0.930(0.04) & \textbf{0.942}(0.03) \\
	    \bottomrule
	    \end{tabular}%
	}
	\label{RQ3-FN}%
\end{table}%

\begin{table}[htbp]
	\centering
	\caption{The Average MCC of ConDF and other feature selection methods using DF classifier}
	\scalebox{0.65}[0.65]{
	    \begin{tabular}{c|ccccccc}
	    \toprule
	    Project & CSDF  & IGDF  & GRDF  & ReFDF & CFSDF & DF    & ConDF \\
	    \midrule
	    Codec & 0.525(0.11) & 0.545(0.09) & 0.554(0.09) & 0.432(0.13) & 0.547(0.10) & 0.476(0.09) & \textbf{0.567}(0.10) \\
	    Colle & 0.723(0.04) & 0.716(0.05) & 0.730(0.07) & 0.683(0.04) & \textbf{0.742}(0.04) & 0.679(0.04) & 0.735(0.04) \\
	    IO    & 0.655(0.07) & 0.666(0.08) & 0.651(0.07) & 0.656(0.09) & 0.651(0.10) & 0.687(0.05) & \textbf{0.712}(0.05) \\
	    Jsoup & 0.421(0.10) & 0.421(0.11) & 0.432(0.10) & \textbf{0.488}(0.12) & 0.471(0.13) & 0.387(0.09) & 0.472(0.14) \\
	    JSqlP & 0.665(0.20) & 0.556(0.32) & 0.609(0.26) & 0.681(0.21) & 0.585(0.28) & \textbf{0.722}(0.14) & 0.685(0.18) \\
	    Mango & 0.686(0.08) & 0.671(0.12) & 0.699(0.09) & 0.354(0.15) & \textbf{0.770}(0.06) & 0.621(0.10) & 0.761(0.10) \\
	    Ormli & 0.731(0.06) & 0.726(0.07) & 0.758(0.06) & 0.741(0.06) & 0.794(0.06) & 0.678(0.05) & \textbf{0.833}(0.04) \\
	    \midrule
	    Average & 0.629(0.11) & 0.614(0.10) & 0.633(0.11) & 0.576(0.14) & 0.651(0.11) & 0.607(0.12) & \textbf{0.681}(0.11) \\
	    \bottomrule
	    \end{tabular}%
	}
	\label{RQ3-MCC}%
\end{table}%

\begin{figure}[htbp]
	\centering
	\subfigure[$\rm{F_{InTrace}}$]{
		\label{fig:RQ3-F}
		\begin{minipage}{0.9\linewidth}
			\centering
			\includegraphics[width=1\linewidth]{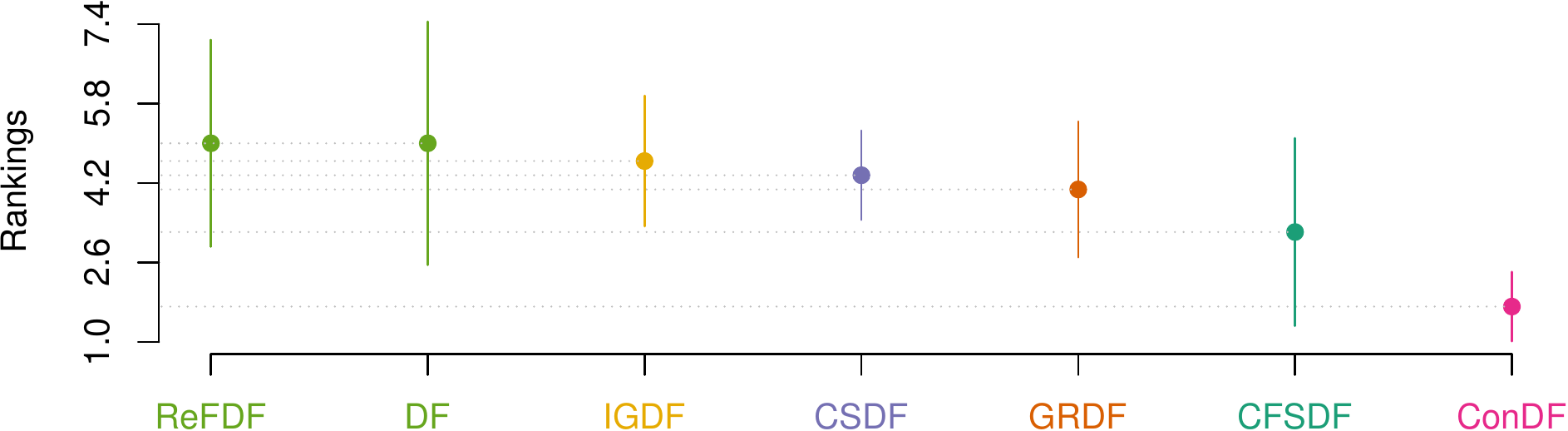}
		\end{minipage}
	}
	\subfigure[$\rm{F_{OutTrace}}$]{
		\label{fig:RQ3-FN}
		\begin{minipage}{0.9\linewidth}
			\centering
			\includegraphics[width=1\linewidth]{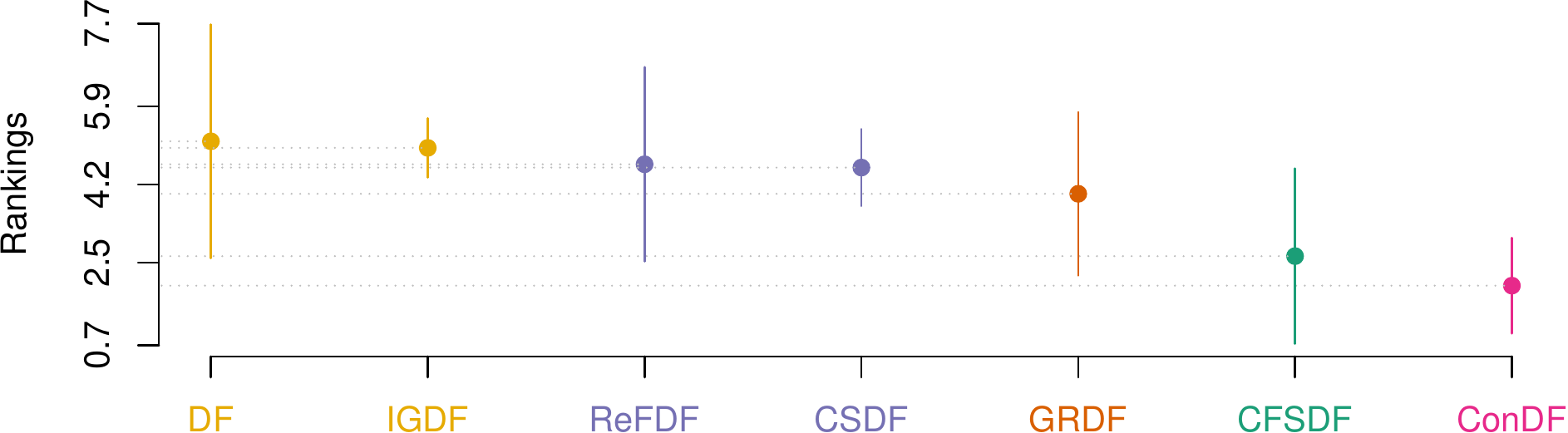}
		\end{minipage}
	}	
	\subfigure[MCC]{
		\label{fig:RQ3-MCC}
		\begin{minipage}{0.9\linewidth}
			\centering
			\includegraphics[width=1\linewidth]{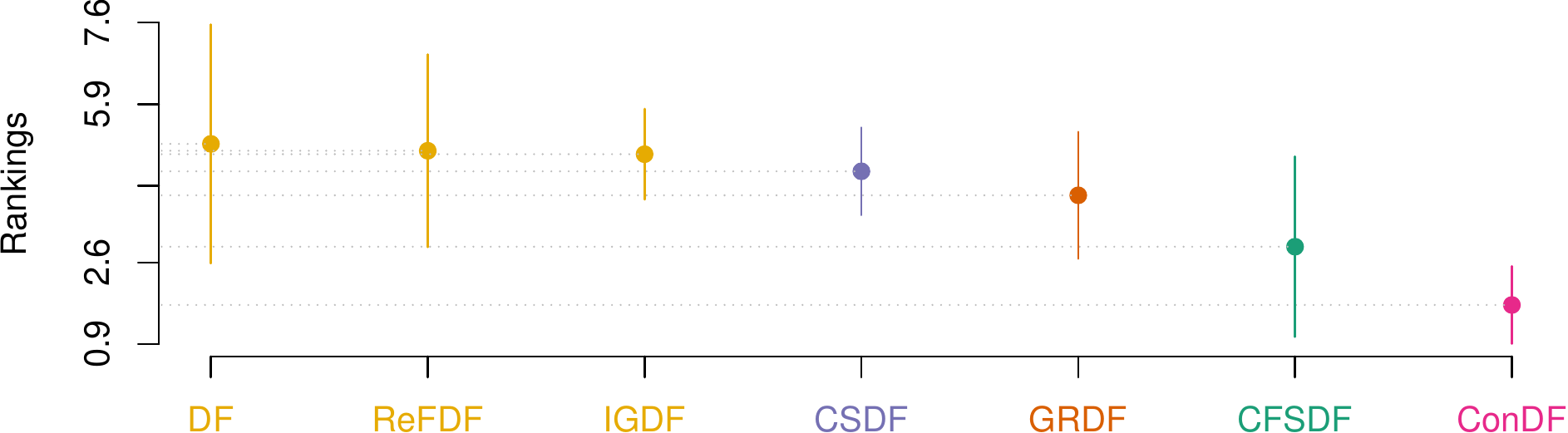}
		\end{minipage}
	}
	\caption{Scott-Knott ESD test for ConDF and other feature selection methods using DF classifier.}
	\label{fig:RQ3-CK-SKESD}
\end{figure}

\subsection{RQ2: How effective our ConDF method is compared with other classifiers with the feature selection method Con?}

\textbf{Motivation}:
Since our crashing fault residence prediction task is a classification problem, the prediction performance is related to the used classifiers. In this work, we apply the DF method as the classifier.
This question is designed to explore whether the feature selection method Con with DF classifier is more effective to achieve better performance than that with other classifiers. 

\textbf{Methods}:
To answer this question, we choose five widely used classifiers, including \textbf{D}ecision \textbf{T}ree (DT), \textbf{S}upport \textbf{V}ector \textbf{M}achine (SVM), \textbf{L}ogistic \textbf{R}egression (LR), \textbf{R}andom \textbf{F}orest (RF), and \textbf{N}earest \textbf{N}eighbor (NN), and combine them with the feature selection method Con as the comparative methods. The five baseline methods are short for ConDT, ConSVM, ConLR, ConRF, and ConNN, individually.

\textbf{Results}:
Table \ref{RQ2-F}, \ref{RQ2-FN}, and \ref{RQ2-MCC} report the results of the indicator values for our ConDF method and the 5 comparative methods in terms of $\rm{F_{InTrace}}$, $\rm{F_{OutTrace}}$, and MCC, individually. 
In terms of $\rm{F_{InTrace}}$, ConDF achieves better performance on 5 out of 7 projects compared with the 5 baseline methods. The average $\rm{F_{InTrace}}$ value by ConDF obtains performance improvements by 7.6\%, 38.6\%, 17.4\%, 179.8\%, and 21.1\% compared with ConDT, ConSVM, ConLR, ConRF, and ConNN, individually. 
In terms of $\rm{F_{OutTrace}}$, ConDF achieves better performance on 6 out of 7 projects compared with the 5 baseline methods. The average $\rm{F_{OutTrace}}$ value by ConDF obtains performance improvements by 2.1\%, 2.7\%, 3.3\%, 4.4\%, and 3.9\% compared with the 5 baseline methods, individually.
In terms of MCC, ConDF achieves better performance on 6 out of 7 projects compared with the 5 baseline methods. The average MCC value by ConDF obtains performance improvements by 13.1\%, 40.4\%, 27.5\%, 147.6\%, and 32.5\% compared with the 5 baseline methods, individually. 
Overall, our proposed ConDF method obtains the best average value and achieves average improvements by 52.9\%, 3.3\%, and 52.2\% in terms of $\rm{F_{InTrace}}$, $\rm{F_{OutTrace}}$, and MCC, individually.

Fig. \ref{fig:RQ2-CK-SKESD} visualizes the corresponding statistical test results for our ConDF method and the 5 baseline methods in terms of all three indicators individually. This figure illustrates that ConDF ranks the first and achieves significantly better performance than all 5 baseline methods in terms of all indicators.

In addition, we also perform the efficiency analysis of the DF model compared with the 5 baseline classifiers. In the experiment, we find that Con runs very fast and its computational time can be almost negligible, thus, we only record the running time of the DF method and the other five classifiers, including the model training and classification time for one data partition. We find that the five classifiers only take a few seconds on all seven projects, while our DF method takes about average 20 seconds among seven projects. The reason that DF spends more time than traditional classifiers is that DF need to construct the cascade structure in which each level consists of many forests. Although the computational time is larger than the traditional classifiers, DF can significantly improve the performance in crashing faults residence prediction task compared with the 5 classic classifiers. In addition, our experiments are conducted on a workstation with Intel(R) Core(TM) i7-4790 CPU with 3.60 GHz $\times$ 4. With the subsequent upgrade of the computer configuration, the execution time of the DF method will be further reduced, and the efficiency of our ConDF method can be further improved. 

\textbf{Answer}: 
The used feature selection method Con with DF classifier significantly outperforms the comparative methods that combine Con with other typical classifiers. It implies that the feature subset selected by Con is more appropriate for DF classifier to obtain better performance for crashing fault residence prediction task.

\subsection{RQ3: Are the selected features by Con more effective for performance improvement than that by other feature selection methods?}

\textbf{Motivation}:
In this work, we use the feature selection method Con, which selects the feature subset with the same discriminability as the original crash data. 
This question is designed to explore whether the features obtained by the Con method are more effective than that by other feature selection techniques to improve the performance of DF classifier for crashing fault residence prediction task. 

\textbf{Methods}: 
To answer this question, we employ five widely used feature selection techniques, including \textbf{C}hi-\textbf{S}quare (CS), \textbf{I}nformation \textbf{G}ain (IG), \textbf{G}ain \textbf{R}atio (GR), \textbf{Re}lief\textbf{F} (ReF), and \textbf{C}orrelation-based \textbf{F}eature \textbf{S}ubset selection (CFS), and combine them with the DF model as the comparative methods. The first four methods are typical feature ranking methods and the last one is a feature subset selection method. For the feature ranking methods, we follow the previous studies \cite{ShivajiWAK09}\cite{shivaji2012reducing} to select the 15\% top ranked features as the candidate which is input into the DF model. For the last method, the number of retained features is determined by itself automatically. The five baseline methods are short for CSDF, IGDF, GRDF, ReFDF, and CFSDF, respectively. 
We select the method that only uses the DF classifier without feature selection as the most basic method for comparison. 

\textbf{Results}:
Table \ref{RQ3-F}, \ref{RQ3-FN}, and \ref{RQ3-MCC} report the results of the indicator values for our ConDF method and the 6 comparative methods in terms of $\rm{F_{InTrace}}$, $\rm{F_{OutTrace}}$, and MCC, individually. 
In terms of $\rm{F_{InTrace}}$, ConDF achieves better performance on 4 out of 7 projects compared with the 6 baseline methods. The average $\rm{F_{InTrace}}$ value by ConDF obtains performance improvements by 7.1\%, 9.6\%, 6.8\%, 16.1\%, 4.3\%, and 10.2\% compared with CSDF, IGDF, GRDF, ReFDF, CFSDF, and DF, individually. 
In terms of $\rm{F_{OutTrace}}$, ConDF achieves better performance on 3 out of 7 projects compared with the 6 baseline methods. The average $\rm{F_{OutTrace}}$ value by ConDF obtains performance improvements by 0.9\%, 0.9\%, 0.6\%, 1.3\%, 0.3\%, and 1.3\% compared with the 6 baseline methods, individually. 
In terms of MCC, ConDF achieves better performance on 3 out of 7 projects compared with the 6 baseline methods. The average MCC value by ConDF obtains performance improvements by 8.3\%, 10.9\%, 7.6\%, 18.2\%, 4.6\%, and 12.2\% compared with the 6 baseline methods, individually. 
Overall, our proposed ConDF method obtains the best average value and achieves average improvements by 9.0\%, 0.9\%, and 10.3\%, in terms of $\rm{F_{InTrace}}$, $\rm{F_{OutTrace}}$, and MCC, individually.

Fig. \ref{fig:RQ3-CK-SKESD} visualizes the corresponding statistical test results for our ConDF method and the 6 baseline methods in terms of all three indicators individually. This figure illustrates that ConDF ranks the first and performs significantly better than all baseline methods in terms of all indicators. 

Moreover, we record the number of features selected by Con. The first four baseline methods reserve 15\% of the original features, i.e., 14 features, for all projects. CFSDF and our proposed ConDF method reserve 6 and 8 features across these projects on average, respectively. From this point of view, our methods can select relatively few but representative features to achieve better performance. In addition, different from the first four feature ranking based methods that measure the relative importance of each feature towards the labels with a statistical value, our Con method is a feature subset selection method that calculates the consistency between a feature subset and labels. It is because that Con is a multivariate measure technique as it measures the merit of a subset of features at a time. 

\textbf{Answer}:
The feature subset selected by the Con method is more effective for the DF classifier to achieve better crashing fault residence prediction performance compared with the feature subsets selected by other classic feature selection methods.

\section{Threats to Validity}\label{sec6}
\subsection{Threats to Internal Validity}
Threats to internal validity are the potential implementation faults of the methods in our experiments. To reduce the threats, we carefully implement the functions that we need based on the deep forest source code provided by the original authors. Meanwhile, we take full advantage of the off-the-shelf implementation by the scikit-learn library and WEKA toolkit to implement the used feature selection method and other comparative methods for minimizing the underlying faults.

\subsection{Threats to External Validity}
Since we conduct experiments on a publicly available benchmark dataset which includes seven open source Java projects, the threats to external validity focus on generalizing our results to the projects developed with other languages. In addition, as the simulative crashes with single-point mutation are generated by using the program mutation testing tool, such generated crashes do not fully reveal the realistic ones. Thus, we need to conduct extra experiments on real-world crashes and with large-scale data to verify the generalization of our results.

\subsection{Threats to Construct Validity}
The threats to construct validity concentrate on the reasonability of the used performance evaluation indicators and statistical test method. We employ 3 commonly-used indicators to evaluate the prediction performance of our ConDF method for crashing fault residence prediction, which makes our assessment more comprehensive. In addition, we apply a state-of-the-art statistic test method, called Scott-Knott ESD test, to conduct the analysis of significant differences between ConDF and other comparative methods,  which makes our evaluation more reliable.

\section{Conclusion}\label{sec7}
In this work, we propose a novel composite method, called ConDF, to predict whether the crashing fault resides inside the stack trace or not. ConDF first employs a consistency-based feature subset selection method to reduce the feature dimension by reserving the important and representative features to replace the original feature set. 
Meanwhile, the simplified version of deep forest method is used to build the classification model. We evaluate the performance of our ConDF method on seven open source software projects with three indicators. The results illustrate that our proposed ConDF method performs significantly better than five ensemble based methods with our feature selection method, five typical classifiers with our feature selection method, six classic feature selection methods (containing a method without feature processing) with DF classifier, and the state-of-the-art method CraTer for predicting the crashing fault residence. 
We release our experimental scripts and the used benchmark dataset at https://github.com/sepine/ConDF.

In the future, we plan to apply our ConDF method to projects developed in other programming languages and with real-world crashes. In addition, we plan to consider the class imbalanced issue of the crash data into the process of model building. 

\section*{Acknowledgment}

This work was supported by the National Natural Science Foundation of China under Grants (No. 61972290), the National Natural Science Foundation of China (No.62002034), the Fundamental Research Funds for the Central Universities (Nos.2020CDCGRJ072 and 2020CDJQY-A021), China Postdoctoral Science Foundation (No.2020M673137), and the Natural Science Foundation of Chongqing in China (No.cstc2020jcyj-bshX0114).



\bibliographystyle{IEEEtran}
\bibliography{mybibfile}

\begin{thebibliography}{10}
\providecommand{\url}[1]{#1}
\csname url@samestyle\endcsname
\providecommand{\newblock}{\relax}
\providecommand{\bibinfo}[2]{#2}
\providecommand{\BIBentrySTDinterwordspacing}{\spaceskip=0pt\relax}
\providecommand{\BIBentryALTinterwordstretchfactor}{4}
\providecommand{\BIBentryALTinterwordspacing}{\spaceskip=\fontdimen2\font plus
\BIBentryALTinterwordstretchfactor\fontdimen3\font minus
  \fontdimen4\font\relax}
\providecommand{\BIBforeignlanguage}[2]{{%
\expandafter\ifx\csname l@#1\endcsname\relax
\typeout{** WARNING: IEEEtran.bst: No hyphenation pattern has been}%
\typeout{** loaded for the language `#1'. Using the pattern for}%
\typeout{** the default language instead.}%
\else
\language=\csname l@#1\endcsname
\fi
#2}}
\providecommand{\BIBdecl}{\relax}
\BIBdecl

\bibitem{wong2016survey}
W.~E. Wong, R.~Gao, Y.~Li, R.~Abreu, and F.~Wotawa, ``A survey on software
  fault localization,'' \emph{IEEE Transactions on Software Engineering (TSE)},
  vol.~42, no.~8, pp. 707--740, 2016.

\bibitem{gu2019does}
Y.~Gu, J.~Xuan, H.~Zhang, L.~Zhang, Q.~Fan, X.~Xie, and T.~Qian, ``Does the
  fault reside in a stack trace? assisting crash localization by predicting
  crashing fault residence,'' \emph{Journal of Systems and Software (JSS)},
  vol. 148, pp. 88--104, 2019.

\bibitem{xu2020imbalanced}
Z.~Xu, K.~Zhao, M.~Yan, P.~Yuan, L.~Xu, Y.~Lei, and X.~Zhang, ``Imbalanced
  metric learning for crashing fault residence prediction,'' \emph{Journal of
  Systems and Software (JSS)}, vol. 170, p. 110763, 2020.

\bibitem{xu2019identify}
Z.~Xu, T.~Zhang, Y.~Zhang, Y.~Tang, J.~Liu, X.~Luo, J.~Keung, and X.~Cui,
  ``Identifying crashing fault residence based on cross project model,'' in
  \emph{Proceedings of the 30th International Symposium on Software Reliability
  Engineering (ISSRE)}.\hskip 1em plus 0.5em minus 0.4em\relax IEEE, 2019.

\bibitem{zhou2017deep}
Z.~Zhou and J.~Feng, ``Deep forest: Towards an alternative to deep neural
  networks,'' in \emph{Proceedings of the 26th International Joint Conference
  on Artificial Intelligence (IJCAI)}, 2017, pp. 3553--3559.

\bibitem{chen2014star}
N.~Chen and S.~Kim, ``Star: Stack trace based automatic crash reproduction via
  symbolic execution,'' \emph{IEEE Transactions on Software Engineering (TSE)},
  vol.~41, no.~2, pp. 198--220, 2014.

\bibitem{nayrolles2015jcharming}
M.~Nayrolles, A.~Hamou{-}Lhadj, S.~Tahar, and A.~Larsson, ``Jcharming: A bug
  reproduction approach using crash traces and directed model checking,'' in
  \emph{Proceedings of the 22nd IEEE International Conference on Software
  Analysis, Evolution, and Reengineering (SANER)}.\hskip 1em plus 0.5em minus
  0.4em\relax IEEE, 2015, pp. 101--110.

\bibitem{nayrolles2017bug}
M.~Nayrolles, A.~Hamou-Lhadj, S.~Tahar, and A.~Larsson, ``A bug reproduction
  approach based on directed model checking and crash traces,'' \emph{Journal
  of Software: Evolution and Process (JSEP)}, vol.~29, no.~3, p. e1789, 2017.

\bibitem{xuan2015crash}
J.~Xuan, X.~Xie, and M.~Monperrus, ``Crash reproduction via test case mutation:
  let existing test cases help,'' in \emph{Proceedings of the 10th Joint
  Meeting on Foundations of Software Engineering (FSE)}, 2015, pp. 910--913.

\bibitem{soltani2017guided}
M.~Soltani, A.~Panichella, and A.~Van~Deursen, ``A guided genetic algorithm for
  automated crash reproduction,'' in \emph{Proceedings of the 39th IEEE/ACM
  International Conference on Software Engineering (ICSE)}.\hskip 1em plus
  0.5em minus 0.4em\relax IEEE, 2017, pp. 209--220.

\bibitem{sabor2019automatic}
K.~K. Sabor, M.~Hamdaqa, and A.~Hamou-Lhadj, ``Automatic prediction of the
  severity of bugs using stack traces and categorical features,''
  \emph{Information and Software Technology (IST)}, p. 106205, 2019.

\bibitem{soltani2020benchmark}
M.~Soltani, P.~Derakhshanfar, X.~Devroey, and A.~Van~Deursen, ``A
  benchmark-based evaluation of search-based crash reproduction,''
  \emph{Empirical Software Engineering (EMSE)}, vol.~25, no.~1, pp. 96--138,
  2020.

\bibitem{wu2014crashlocator}
R.~Wu, H.~Zhang, S.-C. Cheung, and S.~Kim, ``Crashlocator: locating crashing
  faults based on crash stacks,'' in \emph{Proceedings of the 23rd
  International Symposium on Software Testing and Analysis (ISSTA)}, 2014, pp.
  204--214.

\bibitem{wong2014boosting}
C.-P. Wong, Y.~Xiong, H.~Zhang, D.~Hao, L.~Zhang, and H.~Mei, ``Boosting
  bug-report-oriented fault localization with segmentation and stack-trace
  analysis,'' in \emph{Proceedings of the 30th IEEE International Conference on
  Software Maintenance and Evolution (ICSME)}.\hskip 1em plus 0.5em minus
  0.4em\relax IEEE, 2014, pp. 181--190.

\bibitem{moreno2014use}
L.~Moreno, J.~J. Treadway, A.~Marcus, and W.~Shen, ``On the use of stack traces
  to improve text retrieval-based bug localization,'' in \emph{Proceedings of
  the 30th IEEE International Conference on Software Maintenance and Evolution
  (ICSME)}.\hskip 1em plus 0.5em minus 0.4em\relax IEEE, 2014, pp. 151--160.

\bibitem{gong2014locating}
L.~Gong, H.~Zhang, H.~Seo, and S.~Kim, ``Locating crashing faults based on
  crash stack traces,'' \emph{arXiv preprint arXiv:1404.4100}, 2014.

\bibitem{wu2018changelocator}
R.~Wu, M.~Wen, S.-C. Cheung, and H.~Zhang, ``Changelocator: locate
  crash-inducing changes based on crash reports,'' \emph{Empirical Software
  Engineering (EMSE)}, vol.~23, no.~5, pp. 2866--2900, 2018.

\bibitem{liu2014fecar}
S.~Liu, X.~Chen, W.~Liu, J.~Chen, Q.~Gu, and D.~Chen, ``Fecar: A feature
  selection framework for software defect prediction,'' in \emph{Proceedings of
  the 38th IEEE Annual Computer Software and Applications Conference}.\hskip
  1em plus 0.5em minus 0.4em\relax IEEE, 2014, pp. 426--435.

\bibitem{chen2014two}
J.~Chen, S.~Liu, W.~Liu, X.~Chen, Q.~Gu, and D.~Chen, ``A two-stage data
  preprocessing approach for software fault prediction,'' in \emph{Proceedings
  of the 8th International Conference on Software Security and Reliability
  (SERE)}.\hskip 1em plus 0.5em minus 0.4em\relax IEEE, 2014, pp. 20--29.

\bibitem{liu2015fecs}
W.~Liu, S.~Liu, Q.~Gu, X.~Chen, and D.~Chen, ``Fecs: A cluster based feature
  selection method for software fault prediction with noises,'' in
  \emph{Proceedings of the 39th IEEE Annual Computer Software and Applications
  Conference}, vol.~2.\hskip 1em plus 0.5em minus 0.4em\relax IEEE, 2015, pp.
  276--281.

\bibitem{ni2019empirical}
C.~Ni, X.~Chen, F.~Wu, Y.~Shen, and Q.~Gu, ``An empirical study on pareto based
  multi-objective feature selection for software defect prediction,''
  \emph{Journal of Systems and Software (JSS)}, vol. 152, pp. 215--238, 2019.

\bibitem{cui2019novel}
C.~Cui, B.~Liu, and G.~Li, ``A novel feature selection method for software
  fault prediction model,'' in \emph{Proceedings of the 65th Annual Reliability
  and Maintainability Symposium}.\hskip 1em plus 0.5em minus 0.4em\relax IEEE,
  2019, pp. 1--6.

\bibitem{manjula2019deep}
C.~Manjula and L.~Florence, ``Deep neural network based hybrid approach for
  software defect prediction using software metrics,'' \emph{Cluster
  Computing}, vol.~22, no.~4, pp. 9847--9863, 2019.

\bibitem{xu2016impact}
Z.~Xu, J.~Liu, Z.~Yang, G.~An, and X.~Jia, ``The impact of feature selection on
  defect prediction performance: An empirical comparison,'' in
  \emph{Proceedings of the 27th International Symposium on Software Reliability
  Engineering (ISSRE)}.\hskip 1em plus 0.5em minus 0.4em\relax IEEE, 2016, pp.
  309--320.

\bibitem{ghotra2017large}
B.~Ghotra, S.~McIntosh, and A.~E. Hassan, ``A large-scale study of the impact
  of feature selection techniques on defect classification models,'' in
  \emph{Proceedings of the 14th International Conference on Mining Software
  Repositories (MSR)}.\hskip 1em plus 0.5em minus 0.4em\relax IEEE, 2017, pp.
  146--157.

\bibitem{azzeh2008improving}
M.~Azzeh, D.~Neagu, and P.~Cowling, ``Improving analogy software effort
  estimation using fuzzy feature subset selection algorithm,'' in
  \emph{Proceedings of the 4th International Workshop on Predictor Models in
  Software Engineering}, 2008, pp. 71--78.

\bibitem{oliveira2010ga}
A.~L. Oliveira, P.~L. Braga, R.~M. Lima, and M.~L. Corn{\'e}lio, ``Ga-based
  method for feature selection and parameters optimization for machine learning
  regression applied to software effort estimation,'' \emph{Information and
  Software Technology (IST)}, vol.~52, no.~11, pp. 1155--1166, 2010.

\bibitem{shahpar2016improvement}
Z.~Shahpar, V.~Khatibi, A.~Tanavar, and R.~Sarikhani, ``Improvement of effort
  estimation accuracy in software projects using a feature selection
  approach,'' \emph{Journal of Advances in Computer Engineering and
  Technology}, pp. 31--38, 2016.

\bibitem{hosni2017investigating}
M.~Hosni, A.~Idri, and A.~Abran, ``Investigating heterogeneous ensembles with
  filter feature selection for software effort estimation,'' in
  \emph{Proceedings of the 27th International Workshop on Software Measurement
  and 12th International Conference on Software Process and Product
  Measurement}, 2017, pp. 207--220.

\bibitem{liu2019feature}
Q.~Liu, J.~Xiao, and H.~Zhu, ``Feature selection for software effort estimation
  with localized neighborhood mutual information,'' \emph{Cluster Computing},
  vol.~22, no.~3, pp. 6953--6961, 2019.

\bibitem{zhou2019improving}
T.~Zhou, X.~Sun, X.~Xia, B.~Li, and X.~Chen, ``Improving defect prediction with
  deep forest,'' \emph{Information and Software Technology (IST)}, vol. 114,
  pp. 204--216, 2019.

\bibitem{zhengsoftware}
W.~Zheng, S.~Mo, X.~Jin, Y.~Qu, Z.~Xie, and J.~Shuai, ``Software defect
  prediction model based on improved deep forest and autoencoder by forest,''
  in \emph{Proceedings of the 31st International Conference on Software
  Engineering and Knowledge Engineering (SEKE)}, 2019, pp. 419--540.

\bibitem{zhang2016cross}
F.~Zhang, Q.~Zheng, Y.~Zou, and A.~E. Hassan, ``Cross-project defect prediction
  using a connectivity-based unsupervised classifier,'' in \emph{Proceedings of
  the 38th IEEE/ACM International Conference on Software Engineering
  (ICSE)}.\hskip 1em plus 0.5em minus 0.4em\relax IEEE, 2016, pp. 309--320.

\bibitem{liu1996probabilistic}
H.~Liu, R.~Setiono \emph{et~al.}, ``A probabilistic approach to feature
  selection-a filter solution,'' in \emph{ICML}, vol.~96.\hskip 1em plus 0.5em
  minus 0.4em\relax Citeseer, 1996, pp. 319--327.

\bibitem{dash2000consistency}
M.~Dash, H.~Liu, and H.~Motoda, ``Consistency based feature selection,'' in
  \emph{Pacific-Asia Conference on Knowledge Discovery and Data Mining}.\hskip
  1em plus 0.5em minus 0.4em\relax Springer, 2000, pp. 98--109.

\bibitem{tantithamthavorn2016empirical}
C.~Tantithamthavorn, S.~McIntosh, A.~E. Hassan, and K.~Matsumoto, ``An
  empirical comparison of model validation techniques for defect prediction
  models,'' \emph{IEEE Transactions on Software Engineering (TSE)}, vol.~43,
  no.~1, pp. 1--18, 2016.

\bibitem{ShivajiWAK09}
S.~Shivaji, E.~J. Whitehead~Jr, R.~Akella, and S.~Kim, ``Reducing features to
  improve bug prediction,'' in \emph{Proceedings of the 24th IEEE/ACM
  International Conference on Automated Software Engineering (ASE)}.\hskip 1em
  plus 0.5em minus 0.4em\relax IEEE, 2009, pp. 600--604.

\bibitem{shivaji2012reducing}
S.~Shivaji, E.~J. Whitehead, R.~Akella, and S.~Kim, ``Reducing features to
  improve code change-based bug prediction,'' \emph{IEEE Transactions on
  Software Engineering (TSE)}, vol.~39, no.~4, pp. 552--569, 2012.

\end{thebibliography}

\end{document}